\def\etc{{\it etc.}}
\def\etal{{\it et al.}}
\def\ie{{\it i.e.}}
\def\vs{{\it vs.}}
\def\eg{{\it e.g.}}
\def\ibid{{\it ibid.}}

\def\~{{$\tilde{\phantom{a}}$}}

\def\mumu{$\mu^+\mu^-$\ }

\def\cbar{{\overline c}}

\def\nubar{{\overline \nu}}

\documentclass [12pt] {article}
\usepackage{epsfig}
\usepackage{color}

\def\thebibliography#1{\section{References}\markboth
 {REFERENCES}{REFERENCES}\list
 {[\arabic{enumi}]}{\settowidth\labelwidth{[#1]}\leftmargin\labelwidth
 \advance\leftmargin\labelsep
 \usecounter{enumi}}
 \def\newblock{\hskip .11em plus .33em minus -.07em}
 \sloppy
 \sfcode`\.=1000\relax}
\def\upcite#1{\raise6pt\hbox{\scriptsize
\cite{#1}}}
  \def\lsim{\mathrel {\vcenter {\baselineskip 0pt \kern 0pt
    \hbox{$<$} \kern 0pt \hbox{$\sim$} }}}
    \def\gsim{\mathrel {\vcenter {\baselineskip 0pt \kern 0pt
    \hbox{$>$} \kern 0pt \hbox{$\sim$} }}}

\def\s{\\ [-8pt]}  % use in tables for smaller vertical spacing
\def\hline{\noalign{\hrule \vskip2pt}}

\def\|{\ifmmode\Vert\else \char`\|\fi}
\ifx\oldzeta\undefined                          % hasn't been done yet, so 
  \let\oldzeta=\zeta                            % save old definiton
  \def\zzeta{{\raise 2pt\hbox{$\oldzeta$}}}     % make new definition
  \let\zeta=\zzeta                              % and attatch it
\fi

\ifx\oldchi\undefined                           % hasn't been done yet, so 
  \let\oldchi=\chi                              % save old definiton
  \def\cchi{{\raise 2pt\hbox{$\oldchi$}}}       % make new definition
  \let\chi=\cchi                                % and attatch it
\fi

\def\frac#1#2{{#1 \over #2}}

\def\half{\ifinner {\scriptstyle {1 \over 2}}
   \else {1 \over 2} \fi}

\def\ave#1{\left\langle#1\right\rangle} % \ave{stuff} gives <stuff>
\def\abs#1{\left\vert#1\right\vert}	% \abs{stuff} gives |stuff|
\def\simge{\mathrel{%
   \rlap{\raise 0.511ex \hbox{$>$}}{\lower 0.511ex \hbox{$\sim$}}}}
\def\simle{\mathrel{
   \rlap{\raise 0.511ex \hbox{$<$}}{\lower 0.511ex \hbox{$\sim$}}}}

\def\buildchar#1#2#3{{\null\!                   % \null, cancel space
   \mathop#1\limits^{#2}_{#3}                   % #1, #2 above, #3 below
   \!\null}}                                    % cancel space, \null
\def\overcirc#1{\buildchar{#1}{\circ}{}}

\def\slashchar#1{\setbox0=\hbox{$#1$}           % set a box for #1 
   \dimen0=\wd0                                 % and get its size
   \setbox1=\hbox{/} \dimen1=\wd1               % get size of /
   \ifdim\dimen0>\dimen1                        % #1 is bigger
      \rlap{\hbox to \dimen0{\hfil/\hfil}}      % so center / in box
      #1                                        % and print #1
   \else                                        % / is bigger
      \rlap{\hbox to \dimen1{\hfil$#1$\hfil}}   % so center #1
      /                                         % and print /
   \fi}                                         %

\def\subrightarrow#1{%                          % #1 under arrow
  \setbox0=\hbox{%                              % set a box
    $\displaystyle\mathop{}%                    % no mathop
    \limits_{#1}$}%                             % just limits
  \dimen0=\wd0%                                 % get width
  \advance \dimen0 by .5em%                     % add a bit
  \mathrel{%                                    % space like =
    \mathop{\hbox to \dimen0{\rightarrowfill}}% % arrow to width
       \limits_{#1}}}                           % text below

\def\overlay#1#2{\ifmmode%
\setbox0=\hbox{$#1$}%
\setbox1=\hbox to\wd0{\hss$#2$\hss}\else%
\setbox0=\hbox{#1}%
\setbox1=\hbox to\wd0{\hss#2\hss}\fi%
#1\hskip-\wd0\box1 }

\def\pmb#1{\leavevmode\setbox0=\hbox{#1}%
\kern-.02em\copy0\kern-\wd0
\kern.04em\copy0\kern-\wd0
\kern-.02em\raise.04em\box0 }

\def\vereq#1#2{\lower3pt\vbox{\baselineskip1.5pt \lineskip1.5pt
\ialign{$\m@th#1\hfill##\hfil$\crcr#2\crcr\sim\crcr}}}

\def\tensor#1{\protect\@ontopof{#1}{\leftrightarrow}{1.15}\mathord{\box2}}
\def\overstar#1{\protect\@ontopof{#1}{\ast}{1.15}\mathord{\box2}}
\def\overdots#1{\protect\@ontopof{#1}{\cdots}{1.0}\mathord{\box2}}
\def\overcirc#1{\protect\@ontopof{#1}{\circ}{1.2}\mathord{\box2}}
\def\loarrow#1{\protect\@ontopof{#1}{\leftarrow}{1.15}\mathord{\box2}}
\def\roarrow#1{\protect\@ontopof{#1}{\rightarrow}{1.15}\mathord{\box2}}

\def\@ontopof#1#2#3{%
{\mathchoice
{\@@ontopof{#1}{#2}{#3}\displaystyle\scriptstyle}%
{\@@ontopof{#1}{#2}{#3}\textstyle\scriptstyle}%
{\@@ontopof{#1}{#2}{#3}\scriptstyle\scriptscriptstyle}%
{\@@ontopof{#1}{#2}{#3}\scriptscriptstyle\scriptscriptstyle}%
}%
}

\def\@@ontopof#1#2#3#4#5{%
\setbox0=\hbox{$#4#1$}%
\setbox1=\hbox{$#5#2$}%
\setbox2=\hbox{}\ht2=\ht0 \dp2=\dp0 %
\ifdim\wd0>\wd1 %
\setbox1=\hbox to\wd0{\hss\box1\hss}%
\mathord{\rlap{\raise#3\ht0\box1}\box0}%
\else   %
\setbox1=\hbox to.9\wd1{\hss\box1\hss}%
\setbox0=\hbox to\wd1{\hss$#4\relax#1$\hss}%
\mathord{\rlap{\copy0}\raise#3\ht0\box1}%
\fi
}%

\def\lambdabar{\protect\@lambdabar}
\def\@lambdabar{%
\relax
\bgroup
\def\@tempa{\hbox{\raise.73\ht0
\hbox to0pt{\kern.25\wd0\vrule width.5\wd0
height.1pt depth.1pt\hss}\box0}}%
\mathchoice{\setbox0\hbox{$\displaystyle\lambda$}\@tempa}%
{\setbox0\hbox{$\textstyle\lambda$}\@tempa}%
{\setbox0\hbox{$\scriptstyle\lambda$}\@tempa}%
{\setbox0\hbox{$\scriptscriptstyle\lambda$}\@tempa}%
\egroup
}

\def\corresponds{{\lower.2ex\hbox{=}}{\rm\kern-.75em^\triangle}}
\def\succsim{\succ\kern-.9em_\sim\kern.3em}
\def\precsim{\prec\kern-1em_\sim\kern.3em}
\def\slantfrac#1#2{\kern1em^{#1}\kern-.3em/\kern-.1em_{#2}}

\def\bi{\begin{itemize}}
\def\ei{\end{itemize}}

\begin{document}                        

\begin{titlepage}

%\phantom{a}
%\vskip .5in
\centerline{\LARGE\bf Expression of Interest for R\&D towards}
\vskip.2in
\centerline{\LARGE\bf A Neutrino Factory Based on a Storage Ring}
\vskip.2in
\centerline{\Large\bf and}
\vskip.2in
\centerline{\LARGE\bf a Muon Collider}
\vskip.4in
\centerline{\Large Submitted to the National Science Foundation by}
\vskip.2in
\centerline{\Large\bf The Neutrino Factory and Muon Collider Collaboration}
\vskip.2in
\centerline{Edited by K.T.~McDonald for the Collaboration}
\vskip.2in
\centerline{\large (November 7, 1999)}
\vskip.2in
\begin{center}
\includegraphics*[width=6.in]{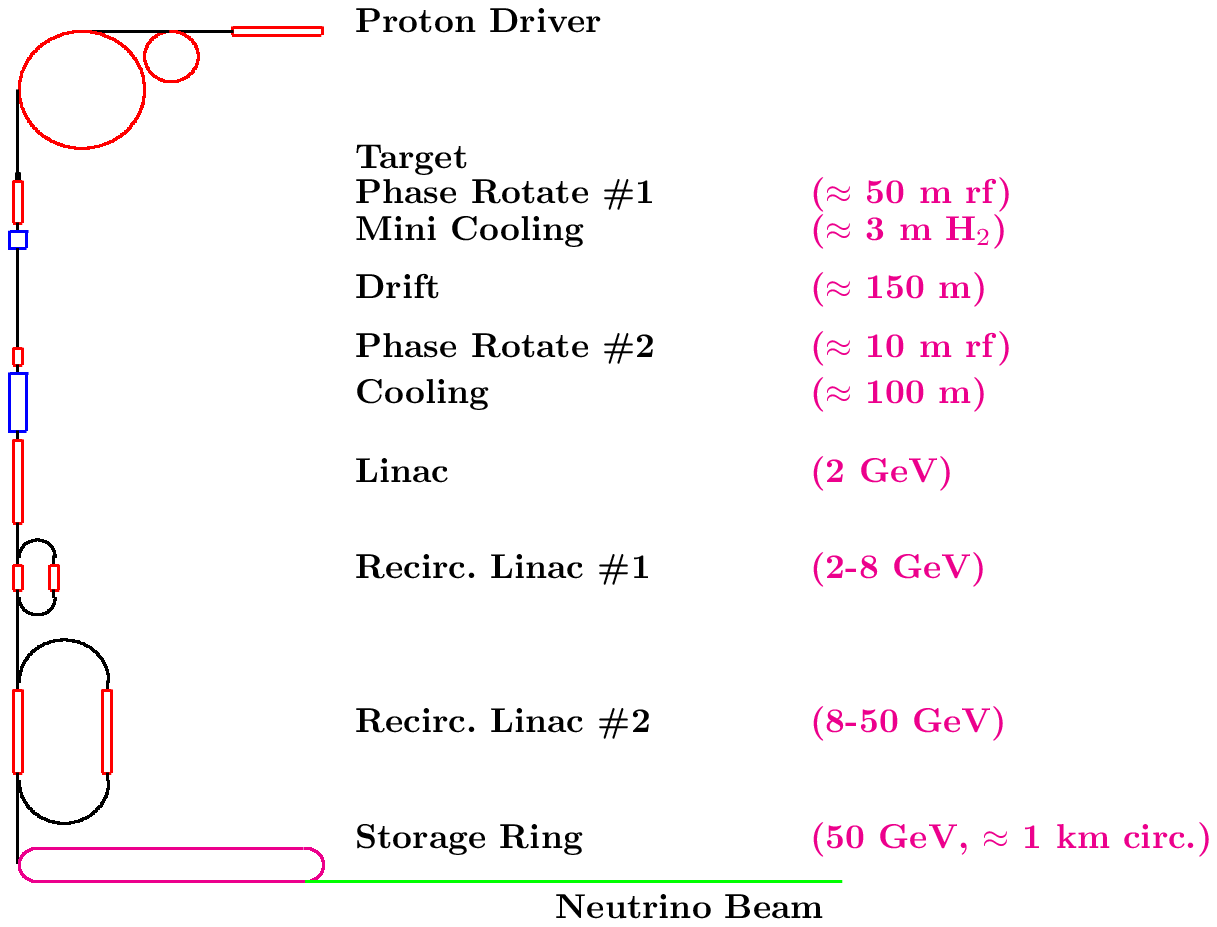}
\vskip.2in
\centerline{Collaboration Home Page: http://www.cap.bnl.gov/mumu/}
\vskip.2in
\centerline{This document resides at 
http://puhep1.princeton.edu/mumu/NSFLetter/nsfmain.ps}
%\vskip-3in
%\centerline{\Huge\bf DRAFT}

\end{center}

\end{titlepage}    

\newpage
\pagestyle{empty}

\centerline{\Large\bf Executive Summary}
\medskip
{\large
Recent evidence from atmospheric, solar, and accelerator neutrinos suggests that
neutrinos have mass, and mix among the flavors $\nu_e$, $\nu_\mu$ and 
$\nu_\tau$.  
%If all experiments are confirmed, there likely exists one or 
%more low-mass sterile neutrinos, $\nu_s$.  
Neutrino mass is evidence for physics
beyond the Standard Model, and has cosmological implications.

%The field of neutrino physics is rich in the variety of experimental techniques
%and in its broad theoretical context.  

Because neutrinos interact so weakly,
unusual efforts are required to detect them.
%and their detection always provides extra-ordinary information.  
Although many of the recent, exciting
results in neutrino physics have been obtained by non-accelerator techniques,
the neutrino mass and mixing parameters appear to be such that a new
generation of accelerator experiments with long baseline distance to the
detectors can perform detailed measurements.
For this, a new source of well-characterized neutrinos is needed.

We are exploring the feasibility of a neutrino factory based on a muon storage
ring.  In this, beams of $\nu_\mu$ and $\nubar_e$ arise from the decay of
$\mu^-$ particles (or alternatively, $\nubar_\mu$ and $\nu_e$ from $\mu^+$). 
The muons come from the decay of low-energy pions produced by a megawatt
proton beam incident on a nuclear target.  The muons are captured into a
magnetic channel, ``cooled" by ionization in liquid hydrogen, 
accelerated to energy of order 50 GeV, and injected into a storage ring.
%where their lifetime is roughly 1000 turns.  
A nonhorizontal ring can
deliver neutrino beams to an on-site detector, as well as to two off-site
detectors separated by global distances.

Such a neutrino factory is a challenging extension of present accelerator
technology.  It is also a natural path to a muon collider, in that both
facilities share many common elements upstream of their storage rings.
Prior to a formal design study, R\&D must be performed in several
keys areas, such detailed simulations and actual targetry and cooling 
experiments.
This in an excellent opportunity to advance the
field of accelerator physics both at national laboratories and at
universities.
}

\newpage

\centerline{\bf The Neutrino Factory and Muon Collider Collaboration}

\begin{flushleft}

%Y.-C.~Chae, 
D.~Ayres,\footnote{Adjunct Member}
M.~Goodman,$^1$
A.~Hassanein,
T.~Joffe-Minor,$^1$
D.~Krakauer,$^1$
J.H.~Norem, 
C.B.~Reed, 
P.~Schoessow,$^1$
D.~Smith, 
R.~Talaga,$^1$
% ...
\\
%\vskip.1in
%International Partners:
%\\
%{\sl CERN}
%\\
%European Study Groups on Muon Machines

\end{flushleft}

\newpage
\pagestyle{plain}
\pagenumbering{roman}
\tableofcontents
\newpage
\listoffigures
\listoftables

\newpage             
\pagenumbering{arabic}

\section{Introduction}

\newcommand{\rmt}{\rm\textstyle}
\newcommand{\rms}{\rm\scriptstyle}
\newcommand{\stw}{\mbox{$\sin^2\theta_W$}}
\newcommand{\nub}{\overline{\nu}}
\newcommand{\nue}{\nu_{e}}
\newcommand{\numu}{\nu_{\mu}}
\newcommand{\nutau}{\nu_{\tau}}
\newcommand{\nx}{\nu_{1}}
\newcommand{\ny}{\nu_{2}}
\newcommand{\nz}{\nu_{3}}
\newcommand{\thetaxy}{\theta_{12}}
\newcommand{\thetayz}{\theta_{23}}
\newcommand{\thetazx}{\theta_{13}}
\newcommand{\Deltaxy}{\Delta m_{12}^{2}}
\newcommand{\Deltayx}{\Delta m_{21}^{2}}
\newcommand{\Deltayz}{\Delta m_{23}^{2}}
\newcommand{\nubmu}{\overline{\nu_{\mu}}}
\newcommand{\nube}{\overline{\nu_{e}}}
\newcommand{\alps}{\mbox{$\alpha_s$}}
\newcommand{\asop}{\mbox{$\frac{\alpha_s}{\pi}$}}
\newcommand{\qsq}{\mbox{$Q^2$}}
\newcommand{\qnsq}{\mbox{$Q_0^2$}}
\newcommand{\mztwo}{\mbox{$M_Z^2$}}
\newcommand{\mz}{\mbox{$M_Z$}}
\newcommand{\mw}{\mbox{$M_W$}}
\newcommand{\mtop}{\mbox{$M_{\rms top}$}}
\newcommand{\mhiggs}{\mbox{$M_{\rms Higgs}$}}
\newcommand{\lmsb}{\mbox{$\Lambda_{\overline{MS}}$}}

%Using muons as a source of neutrinos can be traced back to 25 years ago but 
%it is with the
%recent studies on muon colliders that this idea came back to the 
%front of the stage. It was specifically
%discussed in a FNAL workshop (1997), a BNL workshop (1998) and a study 
%group dedicated to that topic was
%set up at CERN in 1998. 
%$\nu$Fact99 appears as the conclusion of this preparatory work and is the first 
%international workshop on neutrino beams produced by the decay of muons in a 
%storage ring. 
%It was co-sponsored by ICFA and ECFA and attended by 120 participants.
%Proceedings will be published. Physics and accelerator issues were 
%discussed on the same footing. 
%The main physics issue was the theory and detection of neutrino oscillations 
%within the context of three or more families. On the accelerator side, 
%three scenarios have been investigated
%and finally harmonized in a single conceptual design. 
%Last, topics needing R$\&$D are mentioned. 

There is accumulating
evidence for massive neutrinos that mix among flavors. 
The strongest indication is 
the atmospheric neutrino anomaly first observed by
the Kamiokande \cite{Kamiokande} and IMB \cite{IMB} detectors, 
confirmed by the Soudan-2 \cite{Soudan2}
and MACRO \cite{MACRO} detectors, and recently
measured with high statistics by the Super-Kamiokande detector 
\cite{SuperKatmos}.  
In addition, the long-standing deficiency of the solar neutrino
flux measured by the Homestake chlorine experiment \cite{Homestake} is now
supported by data from the Kamiokande \cite{Kamioka}, 
Super-Kamiokande \cite{SuperKsolar}, GALLEX \cite {Gallex}, and SAGE \cite{SAGE}
detectors.  
% An appealing explanation of these data is neutrino oscillations, which, 
% in turn, are due to neutrino masses and mixing.  
These data 
suggest neutrino masses in the range $ \lsim 0.1$ eV for the mass eigenstates
$\nu_i$, $i=1,2,3$ whose linear combinations comprise the neutrinos $\nu_e$,
$\nu_\mu$, and $\nu_\tau$.  Such neutrinos would not be a significant part of
the dark matter of the universe,

The LSND experiment at Los Alamos has 
reported evidence of $\nu_\mu - \nu_e$ oscillations \cite{lsnd}, 
although so far this has
not been confirmed by a similar experiment, KARMEN, at Rutherford 
\cite{KARMEN}.  If confirmed, this results appears to require the existence
of one or more light, sterile neutrinos which could be an
important component of hot, dark matter.

The issue of neutrino mass has spawned a new ``industry'' \cite{nuindustry},
resulting in about three new preprints per day \cite{nuphys}, 
among other activities.  
Excitement is high in the accelerator physics community because
the physics implied by the atmospheric-neutrino results is accessible to
long-baseline accelerator experiments such as K2K \cite{K2K},
Minos \cite{Minos} and NGS \cite{NGS}.
Of course, the LSND experiment was conducted at
a short-baseline accelerator facility, and can be confirmed by future 
accelerator experiments such as MiniBooNE \cite{miniBoone},
ORLanD \cite{Orland}, and CERN P311 \cite{P311}.
Moreover, even the physics associated with many of the
interpretations of the solar-neutrino deficit is accessible to study in
accelerator-based experiments if neutrino-beam fluxes can be
improved by 1-2 orders of magnitude.

To obtain a factor of 100 improvement 
in neutrino flux in a cost-effect manner, a new
approach is called for.  The best prospect appears to be neutrino beams
derived from a muon storage ring, rather than from pion and kaon decay,
although the concept of muon-based neutrino beams needs considerable 
development before it can be realized in the laboratory.

Muon storage rings have been discussed since at least 1960 \cite{Melissring},
and their possible application to neutrino physics was considered as early as
1980 \cite{Cline80}.  
However, storage rings with enough circulating muons
to provide more high-energy neutrinos than from horn beams have been only
recently been considered in the context of muon colliders \cite{status}.
Enthusiasm for muon-based neutrino beams has been fostered by a series of
workshops and studies at Fermilab \cite{mup}, BNL \cite{BNL98}, and
CERN \cite{CERN9902}, resulting in a convergence of international interest
at the NuFact'99 Workshop \cite{nufact99,whitepaper}. 

The neutrino fluxes from these proposed muon-based beams are higher than ever
achieved before, with a better-understood flavor composition, and,
since the neutrino beams from this source would be secondary
beams rather than tertiary beams, they are more collimated than ever
previously imaginable.
%These characteristics have enabled neutrino physicists to
%think about neutrino oscillation measurements in a completely new way.  
Distances between production and detection can now span the
globe, and using the known flavor composition of the beam, one can map out a
plan to measure the neutrino oscillation mixing matrix 
including CP violating effects,  much like
that now underway to study the CKM quark mixing matrix.  

We present a brief review of the physics of neutrino 
oscillations in sec.~2, also including
detector issues most critical for neutrino 
oscillation measurements.  
As an example of how diverse a neutrino program at a storage ring could be, 
highlights of possible nucleon structure and other near-detector 
measurements are given in sec.~3.
The machine itself is discussed in sec.~4, and its possible extension to a
muon collider is considered in sec.~5.
The active theme of this document, research and development towards the
design of a neutrino factory, is discussed in sec.~6.

\section{Neutrino Oscillations} %$S.\;Petcov$}

\subsection{Interpretations of the Data}

The concept of neutrino oscillation was introduced in 1957 \cite{Pontecorvo}
and has been extensively discussed in the literature \cite{oldrev} and
now on the internet \cite{nusites}.
In the example of only two massive neutrinos, with mass eigenstates
$\nu_1$ and $\nu_2$ with mass difference 
$\Delta m$ and mixing angle $\theta$, the flavor eigenstates are
\begin{equation}
\left( \begin{array}{c} \nu_a \\ \nu_b \end{array} \right) =
\left( \begin{array}{cc} \cos\theta & \sin\theta \\
                        -\sin\theta & \cos\theta \end{array} \right) 
\left( \begin{array}{c} \nu_1 \\ \nu_2 \end{array} \right).
\label{k1}
\end{equation}
The probability that a neutrino of flavor $\nu_a$ and energy $E$
 appears as flavor $\nu_b$
after traversing distance $L$ in vacuum is
\begin{equation}
P(\nu_a \to \nu_b) = \sin^2 2\theta \sin^2 \left( {1.27 \Delta m^2 [{\rm eV}^2]
\ L [{\rm km}] \over E [{\rm GeV}] } \right).
\label{k2}
\end{equation}
As the atmospheric neutrino data involves GeV muon neutrinos with
distance scales of the Earth's diameter, this suggests $\Delta m^2$ of order
$10^{-3}$ (eV)$^2$ for $\sin^2 2\theta \approx 1$ \cite{atmosnufits}.  
The solar neutrino data involves MeV electron
neutrinos and distance scales of the radius of the Earth's orbit, suggesting
$\Delta m^2$ of order $10^{-10}$ (eV)$^2$ with $\sin^2 2 \theta \approx 1$
for vacuum oscillations \cite{justso}.
The LSND
result involves 30-MeV muon antineutrino and a distance scale of 30 m,
suggesting $\Delta m^2$ of order 1 (eV)$^2$; large mixing angles are excluded
by reactor data \cite{Bugey}, so $\sin^2 2\theta$ can only be of order
$10^{-2}$ in this case.

Clearly, four different massive neutrinos are required to accommodate all
three results, given their disparate scales of $\Delta m^2$.  The
Standard Model presently includes only three neutrinos with standard
electroweak couplings and $m_\nu < m_Z/2$,
so a ``sterile'' neutrino is required if all the data are correct
\cite{sterile}.  Even discarding
the LSND result, three massive neutrinos are required with a corresponding
$3 \times 3$ mixing matrix (MNS matrix) \cite{MSN}, one of whose
representations is, where c$_{12} = \cos\theta_{12}$, \etc,
\begin{equation}
\left( \begin{array}{c} \nu_e \\ \nu_\mu \\ \nu_\tau \end{array} \right) =
\left( \begin{array}{ccc}
{\rm c}_{12} {\rm c}_{13} & {\rm s}_{12} {\rm c}_{13} & 
        {\rm s}_{13} e^{-i \delta} \\
- {\rm s}_{12} {\rm c}_{23} - {\rm c}_{12} {\rm s}_{13} {\rm s}_{23} e^{i\delta}
& {\rm c}_{12} {\rm c}_{23} - {\rm s}_{12} {\rm s}_{13} {\rm s}_{23} e^{i\delta}
& {\rm c}_{13} {\rm s}_{23} \\ 
{\rm s}_{12} {\rm s}_{23} - {\rm c}_{12} {\rm s}_{13} {\rm c}_{23} e^{i\delta}
& 
- {\rm c}_{12} {\rm s}_{23} - {\rm s}_{12} {\rm s}_{13} {\rm c}_{23} e^{i\delta}
& {\rm c}_{13} {\rm c}_{23} 
\end{array} \right) 
\left( \begin{array}{c} \nu_1 \\ \nu_2 \\ \nu_3\end{array} \right).
\label{k1a}
\end{equation}

In the model of three massive neutrinos, the neutrino oscillation
probabilities of interest depend on six measurable parameters:
three mixing angles 
($\thetaxy$, $\thetazx$, $\thetayz$), and a phase $\delta$ related 
to CP violation as indicated in eq.~(\ref{k1a}); and  two differences of 
the squares of the neutrino masses ($\Deltaxy$ and $\Deltayz$ for instance).
The interpretation of the solar and atmospheric neutrino data
in terms of the three-neutrino oscillation hypothesis suggests 
$|\Deltaxy| \ll |\Deltayz|$, with
$\Deltaxy$ and $\Deltayz$ being responsible 
for the transitions and/or oscillations of 
the solar and atmospheric neutrinos,
respectively.  Then, $|\Delta m^2_{13}| \approx |\Deltayz|$.

The description of the 
atmospheric neutrino data
requires $\Deltayz \approx (2 - 6)\times 10^{-3}~{\rm eV^2}$
and large mixing angle $\thetayz$: 
$\sin^22\theta_{23} \approx (0.9 - 1.0)$.
For $|\Deltaxy|\ll |\Deltayz|$ and with
$\Deltayz$ having a value in the above range,
the nonobservation of oscillations of the
reactor electron antineutrinos
in the CHOOZ experiment \cite{Chooz} 
implies a limit on the angle
$\thetazx$: $\sin^2\theta_{13} < 0.05$.
Given these constraints, the transitions/oscillations
of the solar neutrinos in the three-neutrino
mixing scheme under discussion depend largely on just two parameters:
$\Deltaxy$ and $\sin^22\theta_{12}$.

The presence of matter can strongly modify the
oscillations of electron neutrinos due to their charged-current interaction
(MSW effect \cite{MSW}): in particular,
the oscillations can be resonantly enhanced
by the matter effects even when the oscillation
probabilities are small in vacuum.
This leads to additional interpretations of the solar neutrino data in
which $\Delta m^2_{12}$ can be of order $10^{-5}$ (eV)$^2$ \cite{solarnufits}.
Indeed, there are four presently viable interpretations of the solar neutrino
data:
\begin{itemize}
\item Vacuum oscillation (VO) solution with 
$\Deltaxy \approx (0.5 - 5.0)\times 10^{-10}~{\rm eV^2}$
and $\sin^22\theta_{12} \approx (0.7 - 1.0)$,
\item Low MSW solution corresponding to
$\Deltaxy \approx (0.5 - 2.0)\times 10^{-7}~{\rm eV^2}$
and $\sin^22\theta_{12} \approx (0.9 - 1.0)$,
\item Small mixing angle (SMA) MSW solution with
$\Deltaxy \approx (4.0 - 9.0)\times 10^{-6}~{\rm eV^2}$
and $\sin^22\theta_{12} \approx (0.001 - 0.01)$,
\item Large mixing angle (LMA) MSW solution,
$\Deltaxy \approx (0.2 - 2.0)\times 10^{-4}~{\rm eV^2}$
and $\sin^2\theta_{12} \approx (0.65 - 0.96)$.
\end{itemize}
For the VO, Low, and SMA MSW solutions,
the expressions for the various
transition/oscillation probabilities 
at distances which can be reached on earth 
simplify: they reduce essentially
to the two-neutrino mixing expressions. 
Neglecting the possible matter effects
for simplicity, we can write them in the form
\begin{equation}
P(\nue \to \numu) = \sin^{2}(2 \thetazx) \sin^{2}(\thetayz) \sin^{2}
\left( \frac{1.27 \Deltayz L}{E_{\nu}} \right),
\label{i1}
\end{equation}
\begin{equation}
P(\nue \to \nutau) = \sin^{2}(2 \thetazx) \cos^{2}(\thetayz) \sin^{2}
 \left( \frac{1.27 \Deltayz L}{E_{\nu}} \right),
\label{i2}
\end{equation}
\begin{equation}
P(\numu \to \nutau) = \cos^{4}(\thetazx) \sin^{2}(2 \thetayz) \sin^{2}
\left( \frac{1.27 \Deltayz L}{E_{\nu}} \right).
\label{i3}
\end{equation}
In the case of the large-mixing-angle (LMA)
 MSW solution
there is a known small but non-negligible correction 
in the above expressions  due to the $\Deltaxy$. 

Another type of interpretation is often made of these data, in which the
mass $m$ of a light neutrino is related to an intermediate mass scale
$m_I$ and an heavy mass scale $m_H$ according to the ``seesaw'' mechanism
\cite{seesaw} which predicts
\begin{equation}
m = {m_I^2 \over m_H}.
\label{k3}
\end{equation}
There remains considerable flexibility in the choice of these mass
scales, but a particularly suggestive version \cite{seesaw2} invokes
the vacuum expectation value, 250 GeV, of the Higgs field as the intermediate
mass, so that estimating $m \approx \sqrt{\Delta m^2({\rm atmospheric})} 
\approx 0.06$ eV yields $m_H \approx 5 \times 10^{15}$ GeV.  This scale
is commonly associated with the supersymmetric unification scale in
SO(10) models.  Hence, there is optimism that neutrino mass is evidence that
supersymmetry exists at the GUT scale.   Only a small additional dose of
optimism is required to expect that the supersymmetric partners of known
particles have masses near the intermediate scale, $m_I \approx 250$ GeV,
and will be found during the next decade.

\subsection{The Next Generation of Neutrino Experiments}

With four interpretations of the solar neutrino data, and the two 
interpretations of the LSND data as either right or wrong, there are a total
of eight scenarios for explanations of the data.  The experimental challenge 
is to reduce these to a unique scenario, and to make accurate measurements of
the parameters of that scenario.

It is likely that the next generation of short-baseline accelerator 
neutrino experiments
mentioned previously \cite{miniBoone,Orland,P311} will clarify the status
of the LSND result within 5 years.

Continued operation of Super-Kamiokande, plus the new long-baseline
mentioned previously \cite{K2K,Minos,NGS} will firm up the physics closely
associated with the atmospheric neutrino anomaly over the next decade, but
will have limited ability to explore more than a two-neutrino interpretation.

The solar neutrino spectrum is complex, and all interpretations of the solar
neutrino deficit invoke fortuitous energy dependence in the models.  This
should permit new critical tests of these models as new detectors come into
operation with different energy sensitivities.  

Super-Kamiokande and the SNO experiment \cite{SNO} (which has just
started operation) have good
sensitivity to higher-energy solar neutrinos, whose flux is predicted to
rise with energy in the ``just-so" models.  However, precise interpretation
may be elusive here even with improved statistics, due to uncertainties in
the production rate of $hep$ neutrinos in the Sun.

These high-statistics experiments will also provide more-significant tests of
the dependence of oscillation rates on varying path length (seasonal variation)
implied in the ``just-so'' models,
and on traversal of varying amounts of matter (day/night effect) which
affect some of the MSW solutions.

At the other end of the energy spectrum, the BOREXino liquid scintillator
experiment \cite{Borexino} should be sensitive to the 0.8-MeV $^7$B neutrinos.

Even more ambitious projects, HELLAZ \cite{Hellaz} and HERON \cite{Heron},
plan to use cryogenic techniques to lower their sensitivities
to below the 0.4-MeV maximum of the $pp$ neutrinos whose numbers dominate
the solar neutrino spectrum.

A qualitatively different phenomenon accessible to the SNO experiment is
the comparison of the rates of the reactions $\nu + ^2$H $\to p + p + e$ 
and $\nu + ^2$H $\to p + n + \nu$.  The first reaction can only proceed
via an electron neutrino, while any neutrino flavor can initiate the second.
Hence, if solar electron neutrinos have indeed transformed to other flavors,
the ratio of reaction rates will be less than one.  Such a result will
be unambiguous evidence for neutrino oscillations by itself.

This  extensive program of solar neutrino experiments will certainly greatly
constrain the four present interpretations of the solar neutrino data over
the next decade, although one cannot predict with certainty that only a
single interpretation will then remain.

None of the experiments discussed thus far addresses the long-standing
question of whether neutrinos, if massive, are Dirac neutrinos
(with particles and antiparticles being different: $\nu \neq \nubar$) 
or Majorana neutrinos (with particles and antiparticles the same, as for
photons: $\nu = \nubar$) \cite{Majorana}.  
Theoretically, Majorana neutrinos are more ``natural'', but the question
should be settled experimentally.
This is extremely difficult 
because neutrinos are always (thus far) produced in
weak interactions with a unique helicity, which provides a practical
distinction between neutrino and antineutrino even if there is none in
principle.  Instead, experimental resolution of the question is based on 
the search for neutrinoless double beta decay, (A,Z  $\to$ (A,Z + 2) + 2$e^-$,
which can proceed via annihilation of virtual $\nu_e$ and $\nubar_e$ as
permitted for Majorana, but not for Dirac, neutrinos.  The present 
(model dependent) limit
based on nonobservation of such a decay of $^{76}$Ge is that $m < 0.1$ eV for
Majorana neutrinos.  This limit may be pushed as low as 0.001 eV in the next
decade.

\subsection{The Opportunity for a Neutrino Factory}

Now that there are rough experimental guidelines as to the parameters of
neutrino masses and mixings, one can begin to plan for more extensive studies
than those described in the previous section.  Two prominent features of such
a plan are the need for more neutrinos, and that accelerator experiments with
GeV-energy neutrinos can probe a large fraction of the relevant parameter
space.

The need for more GeV-energy neutrinos leads to a need for GeV proton sources
in the megawatt power range.  Such power sources, when available, could be
used to produce neutrinos via horn beams in the conventional manner.
However, an option has emerged with greater physics flexibility while 
maintaining a comparable or even larger $\nu/p$ ratio than that from horn 
beams.  Namely,
neutrino beams derived from the decay of muons in a storage ring.  Technical 
aspects of muon-based neutrino beams are discussed in sec.~4.
Here, we review the physics opportunities with such beams.

Both $\mu^-$ and $\mu^+$ can be stored in the ring, but only one sign 
will be used at a time.
When, say, $\mu^-$ are stored their decay,
\begin{equation}
\mu^{-} \rightarrow\ e^{-} \: \numu \: \nub_{e},
\label{k4}
\end{equation}
leads to beams that contain nearly equal numbers of $\nu_\mu$ and $\nubar_e$
with spectra that are extremely well known.

At the detectors, the neutrino and the antineutrino may or may 
not have changed their flavor, leading to the appearance of a different flavor
or the disappearance of the initial flavor, respectively.  When detected by a
charged-current interaction, there are 6 classes of signatures in a 
three-neutrino model:
\begin{eqnarray}
& \numu \rightarrow\ \nue \rightarrow e^{-} & \qquad \mbox{(appearance)},
\label{k5a} \\
& \numu \rightarrow\ \numu \rightarrow \mu^{-} & \qquad \mbox{(disappearance)},
\label{k5b} \\
& \numu \rightarrow\ \nutau \rightarrow \tau^{-} &\qquad \mbox{(appearance)},
\label{k5c} \\
& \nub_{e} \rightarrow\ \nub_{e} \rightarrow e^{+}  &\qquad 
\mbox{(disappearance)},
\label{k5d} \\
& \nub_{e} \rightarrow\ \nub_\mu \rightarrow \mu^{+} & \qquad 
\mbox{(appearance)},
\label{k5e} \\
& \nub_{e} \rightarrow\ \nub_\tau \rightarrow \tau^{+} & \qquad 
\mbox{(appearance)}.
\label{k5f}
\end{eqnarray}
A similar list of processes can be written for operation with positive muons. 

Of special interest is process (\ref{k5e}) where a muon of sign different from 
the  parent muon appears.
This is a unique feature of the neutrino factories based on muon beams since 
they are the only 
sources of intense high energy electron (anti)neutrino beams. 

The cases (\ref{k5c}) and (\ref{k5f}) of $\tau$ appearance are only practical
for neutrino beams with 10's of GeV energy.

\subsubsection{Measurements of Masses and Mixing Angles}

First, the high flux of neutrinos coming from the decay ring is ideal to 
measure precisely the
various neutrino cross sections and to explore scenarios with more than
three massive neutrinos, using a compact detector located at a short 
distance. 

By the time a muon storage ring would be built  
it is expected that  two angles $\theta_{23}$ and $\theta_{12}$, 
and the magnitudes of two mass squared differences $\Delta m_{23}^{2}$ and 
$\Delta m_{12}^{2}$
would be known.  This knowledge would come from the solar and atmospheric
neutrino measurements which 
would have been verified by long baseline and reactor experiments, 
for example, MINOS and KamLAND.  
The remaining pieces of the puzzle would be $\theta_{13}$, 
the CP-violating phase $\delta$ and the signs of
the $\Delta m_{ij}^{2}$. In addition, the indicated long-baseline experiments
will not be sensitive to the matter effects in neutrino oscillations
because the distances between the sources and detectors
are not sufficiently large. It would be of fundamental
importance to verify experimentally
the existence of matter effects in neutrino oscillations
by observing directly the modification of the neutrino oscillation
probabilities by these effects.

\begin{figure}[htp]  % h = here, t = top, b = bottom, p = new page
\begin{center}
\includegraphics[width=4in, angle=0, clip]{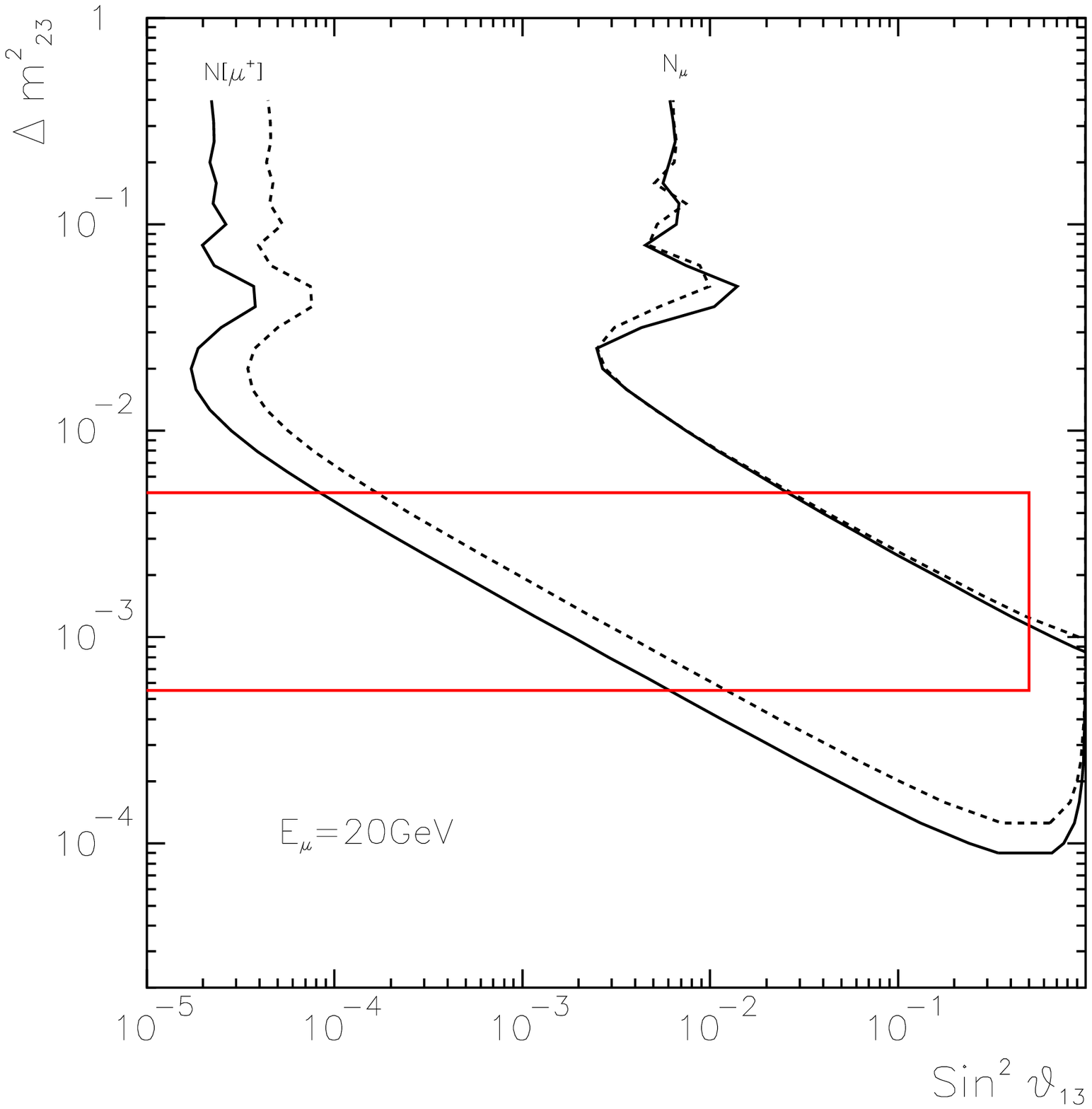}
\parbox{5.5in} % change 5.5in to \hsize for full-width caption
{\caption[ Sensitivity reach in the ($\sin^{2}\thetazx, \Deltayz$) plane.  ]
{\label{figa1} Sensitivity reach in the ($\sin^{2}\thetazx, \Deltayz$) plane
for a 10 kton detector and a neutrino beam from $2 \times 10^{20}$ decays of
20 GeV muons in a storage ring at distance 732 km.
The appearance process $\nubar_e \to \nubar_\mu \to \mu^+$, shown by the lines
on the left, has much greater sensitivity than the disappearance process
$\nu_\mu \to \nu_\mu \to \mu^-$, shown by the lines on the right.
The interior of the box is the approximate region allowed by Super-Kamiokande
data (hep-ph/9811390).
}}
\end{center}
\end{figure}

The third mixing angle
$\thetazx$ can be measured in several channels at a neutrino factory
\cite{nufactstudies}, as can be seen 
from the expressions (\ref{i1})-(\ref{i3}) for various transition 
probabilities.  The detector must be far to avoid
background but not too far ($< 1000$ km) so that the effects of 
$\Deltaxy$ remain negligible and
thus $\delta$ can formally be set to zero. Figure~\ref{figa1} shows the
achievable sensitivity to the yet-unknown value of $\thetazx$.

\subsubsection{Measurement of CP Violation}

The measurement of $\delta$ in a three-neutrino scenario \cite{nufactcp}
%($P.\;Hernandez$, $A.\;Romanino$, $J.\;Sato$) 
relies either on CP violation through the expression
\begin{equation}
A_{\rm CP}=\frac{P(\nue \rightarrow\ \numu)- P(\nub_{e} \rightarrow\ 
\nub_{\mu})}{P(\nue \rightarrow\
 \numu)+ P(\nub_{e} \rightarrow\ \nub_{\mu})},
\label{i4}
\end{equation}
or on time-reversal violation using
\begin{equation}
A_{\rm T}=\frac{P(\nue \rightarrow\ \numu)- P(\numu \rightarrow\ \nue)}
{P(\nue \rightarrow\ \numu)+  P(\numu \rightarrow\ \nue)}.
\label{i5}
\end{equation}
%These asymmetries are indeed larger in the ($\nue$, $\numu$) oscillation. 
The asymmetry (\ref{i4})  can be
measured using wrong-sign muons and the two polarities of the muon beam. 
However, the genuine CP violating
contribution to (\ref{i4}) due to a nonvanishing phase $\delta$ competes with
terms  related to matter effects, \ie, to the different rates of scattering
of $\nue$ and $\nub_{e}$ between source and detector.  The relative strength
of the matter-induced asymmetry increases quadratically with distance, and
dilutes the signal of CP violation in a far detector.  

\begin{figure}[htp]  % h = here, t = top, b = bottom, p = new page
\begin{center}
\includegraphics[width=3in, angle=0, clip]{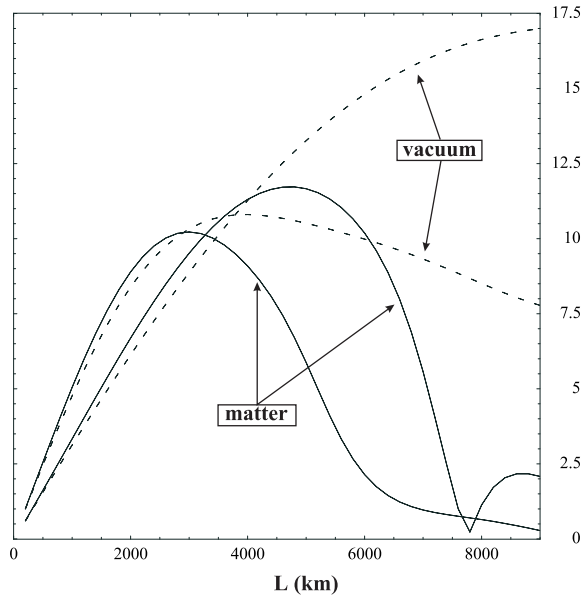}
\parbox{5.5in} % change 5.5in to \hsize for full-width caption
{\caption[ CP violation signal over statistical uncertainties versus distance. ]
{\label{figa2} The CP violating asymmetry (\ref{i4}) divided by statistical 
uncertainties \vs\ distance $L$ for a 10~kton detector in a beam from
$2 \times 10^{21}$ muon decays.  A large angle MSW scenario is supposed,
with $\Delta m^2_{12} = 10^{-4}$ eV$^2$, 
$\Delta m^2_{23} = 2.8 \times 10^{-3}$ eV$^2$,
$\theta_{12} = 22.5^\circ$,
$\theta_{13} = 13^\circ$, 
$\theta_{23} = 45^\circ$, 
and $\delta = -90^\circ$ (corresponding to maximal CP violation).
 The dashed curves ignore matter effects, while
the solid curves include them; the matter effects dominate the asymmetry
for distances beyond 1000 km.  The lower (upper) curves are for $E_\mu = 20$
(50) GeV. From hep-ph/9909254.
}}
\end{center}
\end{figure}

If the solution of the solar neutrinos problem is that 
involving large mixing angles and matter
enhancement (LMA MSW, $\sin^2 2\theta_{12} \approx \sin^2 2\theta_{23} \approx
1$), then there is a possibility of
measuring the CP violating asymmetry (\ref{i4}), whose value is then
\begin{equation}
A_{\rm CP} \approx \abs{{2 \sin\delta \over \sin 2 \theta_{13}}
\sin \left( {1.27 \Delta m^2_{12} L \over E} \right) },
\label{i6}
\end{equation}
provided the detector is 
located sufficiently far and high statistics ($>10^{21}$ 
muons per year) are available.  For all the other solar neutrino solutions
$A_{\rm CP}$ is extremely small, being suppressed by a factor of either
$\sin^2 2\theta_{12}$ or $\Delta m^2_{12}$.  Figure~\ref{figa2}
illustrates the experimental sensitivity to in a large angle MSW scenario.

The asymmetry (\ref{i5}) is not sensitive to matter effects, but relies on 
distinguishing the process $\nu_\mu \to \nu_e \to e^-$ from
$\nubar_e \to \nubar_e \to e^+$.  It will be very difficult to distinguish
electrons from positrons in the detector, but the relative $\nu_\mu$ and
$\nubar_e$ fluxes can be varied by varying the polarization of the muons in the
storage ring \cite{Blondel99}.

%If the three angles in the mixing matrix do not sum to $\pi$ or 
If future experiments confirm the
interpretation of the LSND data that more than three massive neutrinos exist,
%($A.\;Domini$, $C.\;Giunti$, $S.\;Rigolin$), 
then the use of the flavor-rich beams of a neutrino factory is even more 
of an imperative because the parameter space for CP/T violating effects
is considerably enlarged and can 
be successfully explored in experiments with such beams \cite{nufactcp4}.

\subsubsection{Detector Issues} %, $D.A.\;Harris} %^{1}$}

In view of the various experimental signatures (\ref{k5a})-(\ref{k5f}),
an ideal detector would provide identification of both flavor and charge of
all three leptons $e$, $\mu$, and $\tau$.
Muons are the easiest to identify, $\tau$'s are the next easiest if only 
because of their decay to 
muons, and finally electrons are the most difficult.  
Fortunately, there is a very rich program for detectors that only measure
the charge of muons, and hence the oscillation processes (\ref{k5b}) and
(\ref{k5e}) and their conjugates.

\newpage %\medskip

{\bf Baseline Detector Capability} 

\medskip

A  magnetized steel/scintillator sampling calorimeter 
%($F.\;Dydak^{2}$, $J.J.\;Gomez-Cadenas$, $A.\;Cervera$) 
would be one of the far detectors at a muon storage
ring experiment. It could have a hadron energy resolution of 
$0.76/\sqrt{E_{\rm had}[{\rm GeV}]}$,
a hadron angular resolution of $17/\sqrt{E_{\rm had}[{\rm GeV}]}
+ 12/E_{\rm had}[{\rm GeV}]$, 
and much better muon energy and angular resolution.
  
The largest forseeable background in such a detector is charm 
production.  The appearance signal for process (\ref{k5e}) is a ``wrong-sign'' 
muon.  However, if there is enough energy for
charm production in process (\ref{k5b}), the charmed 
particle produced will decay 10\% of the time to a wrong-sign muon 
in the final state.  There is a chance that the associated muon from the 
neutrino interaction vertex is low energy and/or undetected.  
%Secondly, there could
%also be charm produced in neutral-current neutrino scattering,
%which 5\% of the time would produce a wrong-sign muon.  Of these two 
%sources, that from the charged-current process (\ref{k5b}background 
%is expected to be larger.
With kinematic cuts on the muon momentum and its component transverse 
to the hadronic shower, the signal efficiency would be reduced by 25 to 30\%, 
but the backgrounds would be reduced by a factor of $10^{-5}$ to $10^{-6}$
depending on the neutrino energy.
The rejection rate improves faster with energy than does the background,
favoring the use of higher energy muons in the storage ring.

Thus, such a baseline detector would be sufficient for measurements of
$\theta_{13}$ via process (\ref{k5e}),
and the CP-violating phase $\delta$ via the asymmetry (\ref{i4}), both of which
are unlikely to be measured elsewhere and would contribute enormously to the 
field.  

Measurement of the T-violating asymmetry (\ref{i5}) requires separation of
process (\ref{k5a}) from  (\ref{k5d}), ideally performed by measuring the
sign of the electron, and both of these from neutral-current scattering
off electrons.
Depending on the transverse and longitudinal segmentation of the 
scintillator, electron identification is possible, although not 
on an event-by-event basis.  Electron-neutrino charged-current interactions
would be distinguished on average by an energy deposition that was much 
closer to the neutrino interaction vertex, and at an angle 
with respect to the outgoing hadronic shower.  Charge identification 
would not be possible, although from varying the polarization of the 
muon beam one could see how many electron-like events were from $\nub_e$'s, 
and how many were from $\nu_\mu$'s \cite{Blondel99}.

Since a muon-based neutrino factory is a pulsed device with a small duty
factor, cosmic-ray backgrounds will be relatively unimportant.
Hence, there is the option to locate the detectors at
the surface of the Earth, where available infrastructure is more favorable
for very large devices.

Finally, such a baseline detector would have modest detection efficiency for
$\tau$'s via their decay to $\mu$'s, permitting study of process (\ref{k5c})
and (\ref{k5f}) if sufficiently large numbers of neutrinos are available.

\medskip

{\bf Beyond the Baseline Detector} 

\medskip

Additional  technologies must be employed to achieve electron and $\tau$ 
identification and charge measurement on an event-by-event basis.
  
One category of new detectors uses thin ($\sim 100\,\mu$m) sheets of emulsion
combined with thin ($\sim 300\,\mu$m) lead or steel spacers
to  measure kinks that occur 
when a $\tau$ decays.
% by comparing the slope of a track before and after the spacer.  
MINOS 
%($P.\;Lichtfield$, $R.\;Edgecock$) 
is studying the performance of this geometry combined  
with steel for $\tau$ appearance measurements and is likely to install 
such a device if they do see oscillations.  
By comparing the 
change in slope between a few hundred of these sheets, one could make
a 4-$\sigma$ event-by-event measurement on electron or $\tau$ charge.
This technique is practical only in relatively small volumes, and is
perhaps best suited for the near detector, or for the extraordinarily
well collimated neutrino beams from a TeV muon collider.

Detectors which have slightly more promise for 
use on the 10-kton scale  
identify $\tau\to\mu$ decays by their 
difference in kinematics, although they don't see the kink from the decay
itself.  ICARUS, which uses a Liquid Argon TPC detector, has the necessary 
charged track resolution to measure the acoplanarity of an event and 
determine the likelihood of its being a $\tau$ candidate.

\section{Precision High-Rate Neutrino Physics} 

     The advent of a muon storage ring would not only bring about new 
neutrino oscillation measurements, but 
would also usher in a new era for 
high-precision neutrino scattering experiments \cite{nonosc}.
%tests of special and general relativity,
%($C.A.\;Leung$), 
%stopped muons,
%($Y.\;Kuno^{1}$) 
For example, with a detector located 30~m from a 150~m straight
section of a 50-GeV, $10^{21}$-$\mu$/yr muon storage ring, 
the event rate is 40 million events per kilogram per year over a 10 cm 
radius.
%($K.\;Mc\;Farland$). 

To assist in the interpretation of oscillation-related measurements, 
precision measurements would be made  of the total neutrino and antineutrino 
cross sections, as well as of the beam divergence.   

The neutrinos would also be used as precision probes of 
nuclear and nucleon structure, providing additional information to that
obtained in related study using charged lepton beams.
As is well known, neutrino scattering allows a clean separation of the
valence and sea quark distributions, and use of a polarized target permits
characterization of the spin dependence of these distributions.
The near detector is thus the natural successor to 
nucleon structure measurements now underway at HERA, HERMES, Jefferson Lab,
RHIC and elsewhere.

Combined analysis of the scattering of the four neutrino types
$\nu_\mu$, $\nubar_\mu$, $\nu_e$, and $\nubar_e$ off electrons should permit
measurement of the Weinberg angle ten times better than presently known.

A high-flux multi-GeV neutrino beam is also a charm factory, in which
a $\nu_\mu$ beam leads only to $c$ quarks that are tagged by a final-state
$\mu^-$ $(\nu_\mu d \to \mu^- c)$, while $ \nubar_\mu$ beam leads only to
tagged $\cbar$ quarks.
For the beam parameters described above, there would be $10^7$ leptonic 
tagged charm decays 
in only 40 kg-years (not kton-years!), permitting measurements of
$V_{cd}$ to fraction of a percent, and  perhaps even direct observation of
$D^0-{\overline D}^0$ mixing.

\section{A Neutrino Factory}

%%%%%%%%%%%%%%%%%%%%%%%%%%%%%%%%%%%%%%%%%%%%%%%
%Palmer   http://pubweb.bnl.gov/people/palmer/nu/params.ps
%Geer:    http://www.cap.bnl.gov/mumu/status_report.html
%                clic on hep-ph/9906487  for physics and rates
%Blondel  http://alephwww.cern.ch/~bdl/muon/nufacpol.ps
%%%%%%%%%%%%%%%%%%%%%%%%%%%%%%%%%%%%%%%%%%%%%%%%
Relatively complete sketches of a neutrino factory based on a muon storage
ring have emerged only recently via a convergence of earlier visions during 
the NuFact'99 workshop \cite{nufact99,whitepaper}.
Here, we present recent scenarios that consider BNL and FNAL sites as examples
\cite{params}, but note that the conceptual details of a neutrino 
factory are evolving rapidly.

\subsection{Introduction}

Conventional neutrino beams employ a proton beam on a target to 
generate pions, which are focused and allowed to decay into 
neutrinos and, incidentally, muons \cite{MinosTDR}. The muons are discarded 
(stopped in shielding) and the neutrinos ($\nu_\mu$) are 
directed to the detector. In a neutrino factory, pions are made 
the same way and allowed to decay, but it is the decay muons that 
are captured and used. The initial neutrinos from pion decay are discarded, or 
used in a parasitic low-energy neutrino experiment. The muons 
are accelerated and allowed to decay in a storage ring with long 
straight sections. It is the neutrinos from the decaying muons 
(both $\nu_\mu$ and $\nubar_e$) that are directed 
to the detectors.

\begin{figure}[htp]  % h = here, t = top, b = bottom, p = new page
\begin{center}
\includegraphics*[width=4in]{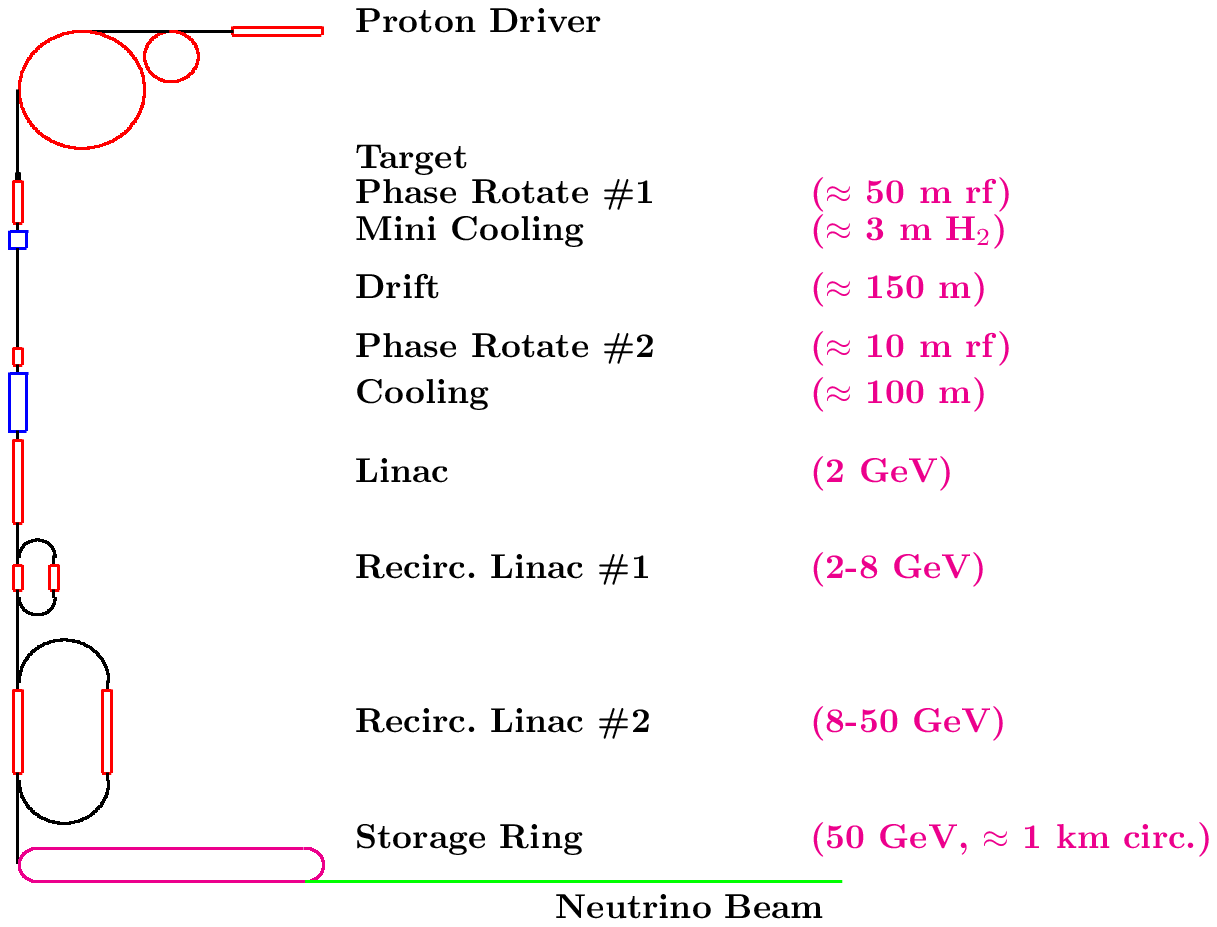}
\parbox{5.5in} % change 5.5in to \hsize for full-width caption
{\caption[ Overview of a neutrino factory. ]
{\label{nufact}  Overview of a neutrino factory based on a muon storage ring.
}}
\end{center}
\end{figure}

The main components of the scenario described here are shown in 
Fig.~\ref{nufact}, and are:
\bi
\item A proton driver of moderate energy ($<50$ GeV) and high average 
power (1-4 MW) similar to that needed for a muon collider, but with 
less stringent requirements on the charge per bunch and somewhat less 
need for power.
\item
A target and pion capture system that can be identical to that for a 
muon collider.
\item
Reduction of the muon energy spread at the expense of 
spreading them out over a longer time interval (longitudinal phase rotation).
The system can be 
designed to correlate the muon polarization with time, allowing 
control of the relative intensity of $\nu_\mu$ and $\nubar_e$ in a forward
beam.  
All this could probably be identical to that for a muon collider.
\item
A limited amount of cooling: about a factor of 50 in six phase-space 
dimensions, 
compared with the factor of $10^6$ needed for a muon collider.
\item
Fast muon acceleration to 50 GeV in a system of an induction linac and two 
recirculating linear accelerators (RLA's). 
This could probably be identical to that for 
a muon collider designed for Higgs production (Higgs Factory).
\item
A collider ring with long straight sections that could point to one or 
more distant neutrino detectors for oscillation studies, and to one or 
more near detectors for high intensity studies.  This ring is rather 
different from one that maximizes luminosity of muon-muon collisions.
\ei

Advantages of a neutrino factory are:
\bi
\item
The spectrum of the neutrinos from muon
decay are very well defined, particularly compared
to conventional neutrino beams from pion decay where 
proton beam size and position, horn current and timing, and the condition of 
the target and horn can all affect the fluxes and backgrounds.
\item
There are almost equal electron and muon neutrino types made, and both 
neutrinos and antineutrinos can be obtained. In beams from pion decay, only 
6muon neutrinos are available with small backgrounds of the other types.
\item
The numbers of neutrinos per initial proton are comparable in the two 
schemes, and for low energy neutrinos there is no flux advantage in the 
factory. But for high energy neutrinos, the conventional approach 
requires high energy protons, of which, for a given power, there will be 
fewer. The neutrino factory can, in principle,  use the same relatively 
low energy protons to produce the same number of neutrinos at any energy
independent of the neutrino energies, and the number can remain high.  
For 50 GeV neutrinos, the gain is between one and two orders of magnitude
over conventional beams.
\item
The intensities are sufficiently high that one can use oscillation 
baselines of the order of the Earth's diameter. One could
build a neutrino factory in the US and detect neutrino oscillations 
in the Gran Sasso detector in Italy, or build the factory in Europe and
direct a beam to the US.   Such intensities and distances 
also allow the study of the neutrino-matter interaction (MSW effect). 
Measurements at multiple distances would, in principle, allow the 
complete determination of the neutrino mass matrix (the equivalent 
of the CKM matrix), including CP violations, while also addressing 
the possible existence of sterile neutrinos.
\item
A neutrino factory is also a first step towards a muon collider.
It would be simpler build than a muon collider, 
would demonstrate most of the components of a collider, and might
be upgradable to a collider.
\ei 

In the remainder of this section, we discuss the various components of a
neutrino factory in greater detail.

\subsection{Proton Driver}

The number of pions per proton produced with an optimized system varies
linearly with the proton energy \cite{muc0061}.
Thus, the number of pions, and the number of muons into which they decay, 
is proportional to the proton beam power. This might suggest that the 
proton energy could be selected arbitrarily, but the situation is more 
complicated.

The total six-dimensional emittance of the produced muons depends on, 
among other things, the pion bunch length, and thus on the rms proton bunch 
length $\sigma_p$ if that length is longer than a length $c~\tau_{\rm decay}$ 
that is characteristic to the decay process:
\begin{equation}
\tau_{\rm decay}  = {(m_\pi - m_\mu)\over m_\pi}~ 
{1\over \gamma_\pi^2}~\tau_\pi,
\label{p1}
\end{equation}
where $\tau_\pi$ is the pion lifetime and $\gamma_\pi m_\pi$ is the 
pion energy. The pion yield peaks at $E_\pi \approx 300$ MeV, which gives 
$\tau_{\rm decay} \approx 1$ nsec. 
This, if the proton energy is low, can imply a large tune shift in the 
proton ring prior to extraction:
\begin{equation}
\Delta\nu~ \propto~ {n_p~ C \over \sigma_t~\epsilon_\perp~\gamma_p^2}
~\propto~{n_p \over \ave{B}~ \sigma_t~\epsilon_\perp~\gamma_p},
\label{p2}
\end{equation}
where $n_p$ is the number of protons in a bunch,
$C$ is the circumference of the proton driver, 
$\ave{B}$ is the average bending field, and 
$\epsilon_\perp$ is the transverse emittance of the protons. The above 
dependency favors a higher proton energy. 

It also favors a high repetition rate with relatively fewer protons
 per bunch, but once again the situation is complicated. The total 
six-dimensional emittance of the produced pions depends also on the
number of proton bunches employed to fill the storage ring. This favors 
a small number of large proton bunches in the driver, and thus a larger 
tune shift. 

However, a high driver repetition rate with smaller numbers of protons 
per fill would not increase the emittance per fill and would still 
reduce the tune shift. The difficulty with this approach is that the 
higher repetition rate increases the wall power required for the 
pulsed rf needed for acceleration and cooling.

These considerations favor a proton driver of 15-25 GeV energy, 1-4 MW
power, with a ring cycling at 5-15 Hz, and a bunch length of order 1 nsec.
Each cycle accelerates about $10^{14}$ protons in 4-6 bunches space about
150~m apart.  Such a
proton driver has significantly higher power than any in present use in the
high energy community, and is comparable to those under design for
neutron spallation sources.

\subsection{Target and Capture}

To maximize the muon yield from pion decay, pions are captured from the
peak of their production spectrum at around 300 MeV/$c$ longitudinal momentum.
The corresponding transverse momenta extend to beyond 200 MeV/$c$, so a
rather diffuse cloud of pions must be captured.  This is best done with a
solenoidal magnetic field, whose acceptance of particles at large angles
is much superior to that of a sequence of quadrupoles.  Indeed, solenoid
magnets must be used to contain the pion/muon beam over much of its length.
The target is surrounded by a 20-T hybrid solenoid magnet \cite{nhmfl-lhfs},
followed by an adiabatic transition to the 1-T field of the decay and
phase rotation channel.

%The yield of pion/muons per unit of proton beam power is largely independent
%of the proton beam energy, although the yield of negative particles decreases
%for $E_p \lsim 2$ GeV.  

The large pulse of energy deposited by the 1-4 MW proton beam in the target 
on nsec time scales lead to
transient pressure waves that are problematic for the long-term survival of
solid targets.  Therefore, a target based on a free mercury jet is under
serious study \cite{muc0061}, with a moving belt target as a backup 
concept \cite{bandsaw}.  At lower beam powers, a radiatively cooled carbon
target may be viable.  However, the yield of pions per proton is higher for a 
high-$Z$ target material.

\begin{figure}[htp]  % h = here, t = top, b = bottom, p = new page
\begin{center}
\includegraphics[width=4in, angle=0, clip]{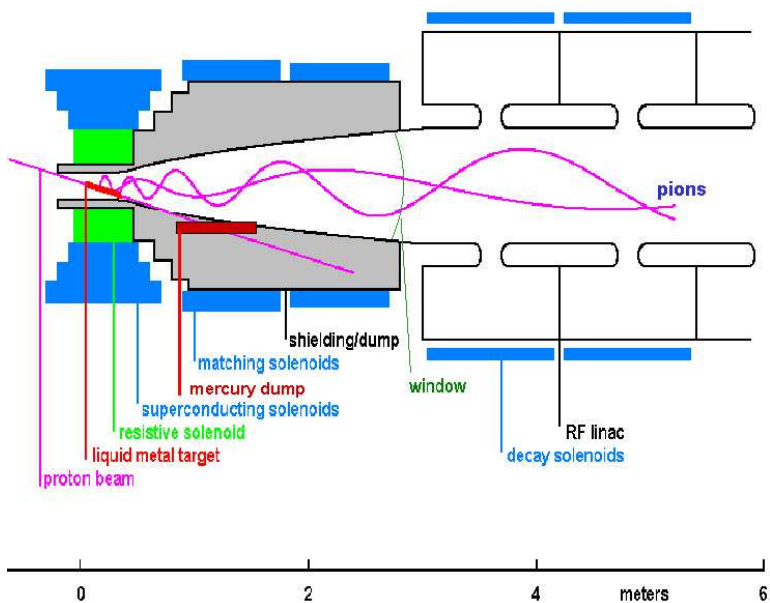}
\parbox{5.5in} % change 5.5in to \hsize for full-width caption
{\caption[ Targetry, pion capture, and beginning of phase rotation. ]
{\label{capt} Targetry, pion capture, and beginning of phase rotation.
}}
\end{center}
\end{figure}

The target and proton beam are at an angle to the axis of the capture
system to minimize re-absorption of the spiralling pions in the target,
and to permit dumping of the proton beam to the side of the system, perhaps
in a pool of mercury.  Figure~\ref{capt} sketches the main features of the
target and capture apparatus, along with the beginning of the phase rotation
channel.

The capture system is very similar to that considered for a 
muon collider source \cite{status}.  

\subsection{Phase Rotation \#1}

An early, high-gradient phase rotation is required if muon polarization 
is to be selected without particle loss. Forward decays, having 
one polarization, yield higher energy muons than backward decays, 
which have the other. If full phase rotation occurred before decay,
then polarization and final energy are fully correlated, but
 significant correlation is obtained even with partial rotation
before decay. The essential requirement is that significant energy 
changes occur before the decay. Phase rotation after decay cannot 
distinguish energy changes due to decay kinematics from the energy 
spread of the initial pions, so there is no way to 
separate the different polarizations. 

\begin{figure}[htp]  % h = here, t = top, b = bottom, p = new page
\begin{center}
\includegraphics[width=3.5in, angle=0, clip]{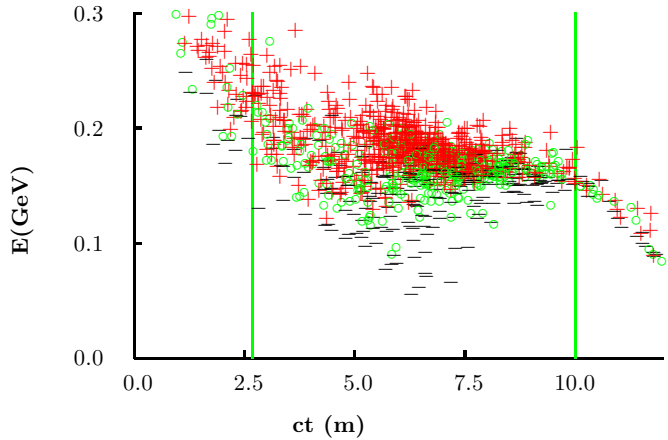}
\parbox{5.5in} % change 5.5in to \hsize for full-width caption
{\caption[ The muon bunch at the end of the first phase rotation. ]
{\label{sim1} The longitudinal-phase-space distribution of the muon bunch at 
the end of the first phase rotation.  
Color and symbols indicate polarization $P$: 
+ (red): $P > 0.3$, o (green): $0.3 > P > -0.3$, $-$ (black): $-0.3 < P$.
}}
\end{center}
\end{figure}

The first phase rotation is accomplished by a sequence of low-frequency rf
cavities that reside inside a solenoid magnet which contains the beam
transversely.  The first cells of this are sketched in Fig.~\ref{capt}.
At the end of this first phase rotation stage, the bunch length has increased 
by a factor of 6 and the energy spread has decreased by the same amount.
Figure~\ref{sim1} shows a simulation of the bunch at the end of the first
phase rotation.

Alternative scenarios without this first stage of phase rotation are under
study \cite{muc0052-59}, always with the result that the polarization 
separation will be lost.

\subsection{Mini Cooling}

Reduction of the phase volume of the muon beam must be accomplished before
the muons decay, which limits the applicability of stochastic cooling and
electron cooling.  Rather, we propose to use
the technique of ionization cooling
\cite{ioncool}    %Oneill,mura126,Kolomensky,Budker67,Ado,Balbekov96}
in which the muons lose both transverse and longitudinal momentum while
passing through bulk matter, and only longitudinal momentum is restored
via rf acceleration.
This technique is uniquely applicable to muons because of
their minimal interaction with matter, and can be performed in less than a
microsecond.

The first stage of cooling at a neutrino factory, called
mini cooling, consists simply of a hydrogen absorber in a solenoidal 
field, and serves two purposes. It reduces the muon energies so that the 
subsequent drift length for a second phase rotation could be kept 
short. It also lowers the transverse emittance by almost a factor 
of two. 

\begin{figure}[htp]  % h = here, t = top, b = bottom, p = new page
\begin{center}
\includegraphics[width=3.5in, angle=0, clip]{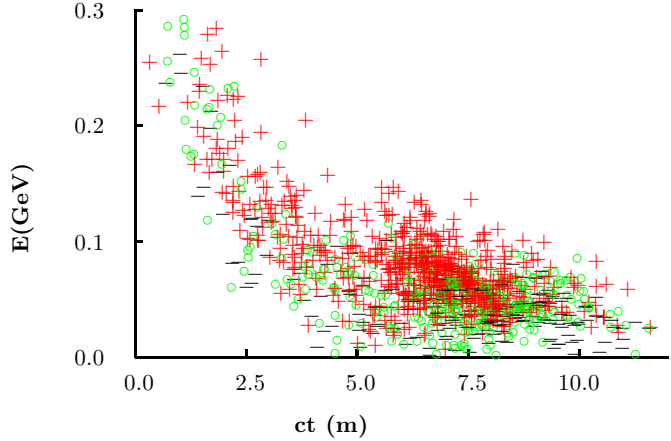}
\parbox{5.5in} % change 5.5in to \hsize for full-width caption
{\caption[ The muon bunch after mini cooling. ]
{\label{sim1a} The longitudinal-phase-space distribution of the muon bunch after
the mini cooling by liquid hydrogen.
Color and symbols indicate polarization $P$: 
+ (red): $P > 0.3$, o (green): $0.3 > P > -0.3$, $-$ (black): $-0.3 < P$.
}}
\end{center}
\end{figure}

In a current simulation \cite{params}, the mini cooling was done in a single 
hydrogen absorber placed in a fixed magnetic field of 1.25~T, with
simulated results as shown in Fig.~\ref{sim1a}.  Such 
cooling introduces canonical angular momentum and it will probably 
be desirable to do the mini cooling in two stages with a field 
reversal between them. 

\subsection{Phase Rotation \#2}

The purpose of phase rotation is to minimize the muon momentum spread,
which can be done at the expense of 
lengthening the bunch up to a distance approaching the initial 
proton bunch spacing ($\approx$ 150 m in the example discussed here). 
The very long 
resulting bunch is then rebunched at a higher frequency ($\approx$ 175 MHz), 
yielding a train of about 30 individual muon bunches for every 
initial proton one.

In addition, this phase rotation results in the polarization being 
correlated with time, \ie, bunch number, instead of energy. This
correlation can, in principle, be preserved thereafter.

The second phase rotation is performed by a drift ($\approx 150$ m), 
followed by energy correction, followed by bunching.

In the present example, an induction linac ($\approx 100$ MeV acceleration) 
is used in which the pulse shape is tailored to correct the time-energy 
correlation generated by the drift. The induction linac must 
supply a rapid train of accelerations, spaced by the proton 
bunch spacing, and equal in number to the number of proton 
bunches.

\begin{figure}[htp]  % h = here, t = top, b = bottom, p = new page
\begin{center}
\includegraphics[width=3.5in, angle=0, clip]{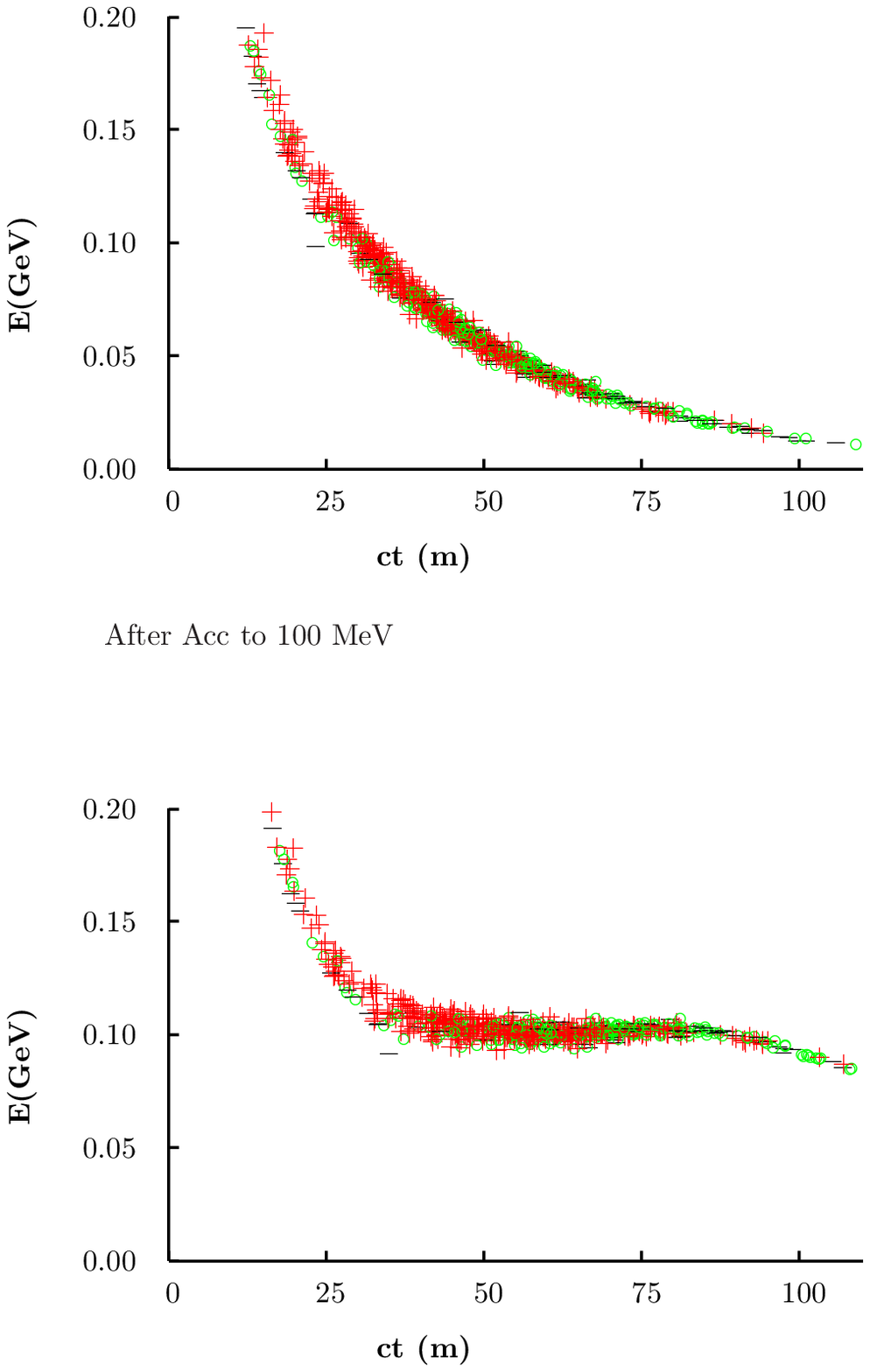}
\parbox{5.5in} % change 5.5in to \hsize for full-width caption
{\caption[ The muon bunch before and after the induction linac. ]
{\label{sim2} The longitudinal-phase-space distribution of the muon bunch after
the second phase rotation (top), and after the induction linac (bottom).
Color and symbols indicate polarization $P$: 
+ (red): $P > 0.3$, o (green): $0.3 > P > -0.3$, $-$ (black): $-0.3 < P$.
}}
\end{center}
\end{figure}

\begin{figure}[htp]  % h = here, t = top, b = bottom, p = new page
\begin{center}
\includegraphics[width=3.5in, angle=0, clip]{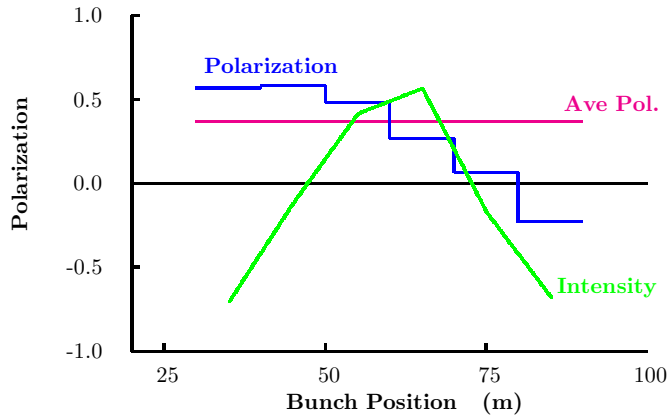}
\parbox{5.5in} % change 5.5in to \hsize for full-width caption
{\caption[ The muon polarization after the induction linac. ]
{\label{pol1} The muon polarization and intensity as a function of position
in the bunch train after the induction linac.
}}
\end{center}
\end{figure}

\begin{figure}[htp]  % h = here, t = top, b = bottom, p = new page
\begin{center}
\includegraphics[width=3.5in, angle=0, clip]{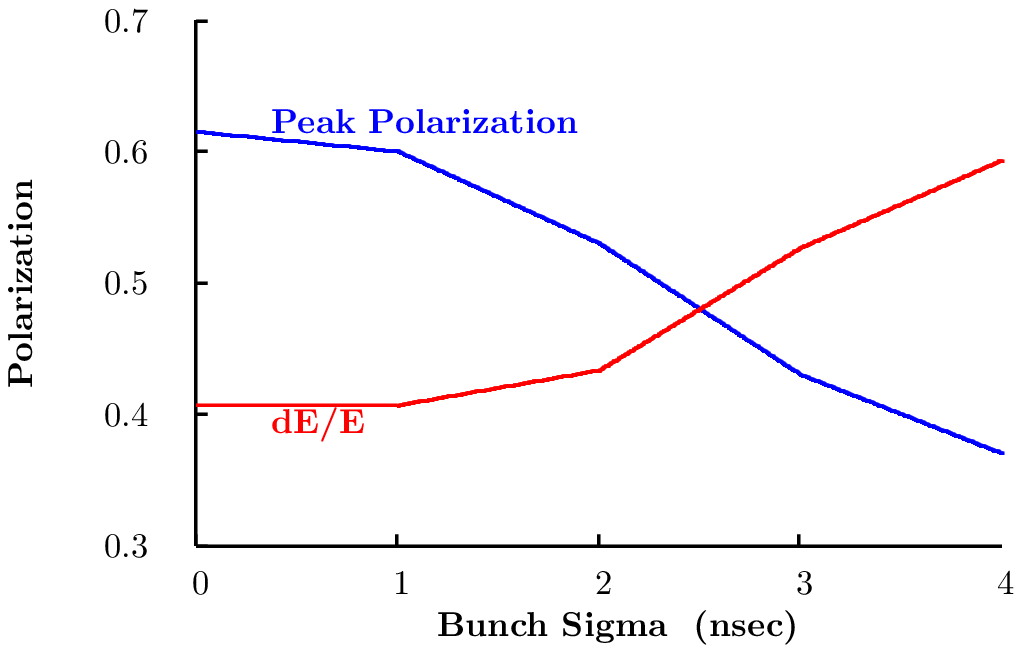}
\parbox{5.5in} % change 5.5in to \hsize for full-width caption
{\caption[ The muon polarization \vs\ proton bunch length. ]
{\label{pol2} The muon polarization after the induction linac as a function of 
the proton driver  bunch length.
}}
\end{center}
\end{figure}

Figure \ref{sim2} shows the simulated  energy \vs\ time distributions 
after the drift, and and after the energy correction in the induction linac.
The simulated final polarizations \vs\ bunch position are shown in 
Fig.~\ref{pol1}.
The maximum muon polarization is a rapid function of the initial proton 
bunch length, as shown in Fig.~\ref{pol2}.
In the simulation, the average muon polarization at the end of the induction
linac is 0.37, and the momentum spread is $dp/p \approx  2$ \%.  
If only 20\% of the muons are kept, the polarization could be 0.6.

Bunching can be done either before or after the energy correction. 
The bunching frequency considered here is a multiple of 350 MHZ, 
the frequency of the superconducting cavities that are assumed to be 
used in the later acceleration. 

More efficient bunching may be possible if the initial energy 
is lower and the bunching is done together with acceleration
\cite{muc0052-59}.
This suggests that a second mini cooling with about 1~m of hydrogen 
could be used to lower the muon energies to about 25 MeV, followed 
by the bunching and acceleration back to 100 MeV. 
The addition of the second mini cooling would further reduce the 
required conventional cooling to follow \cite{muc0055}.

\subsection{Cooling}

A simple comparison of the total produced six-dimensional emittance 
and the total acceptance of a plausible storage ring indicates 
that cooling should not be needed. But without cooling, the muon 
accelerator would have to have a transverse rms acceptance of $\approx 20$
$\pi$ mm-rad (full acceptance $\approx$ 0.2 $\pi$ m-rad). This 
we have shown is possible with large-aperture solenoid focusing 
and low-frequency rf, but would be expensive.

\begin{figure}[htp]  % h = here, t = top, b = bottom, p = new page
\begin{center}
\includegraphics[width=3.5in, angle=0, clip]{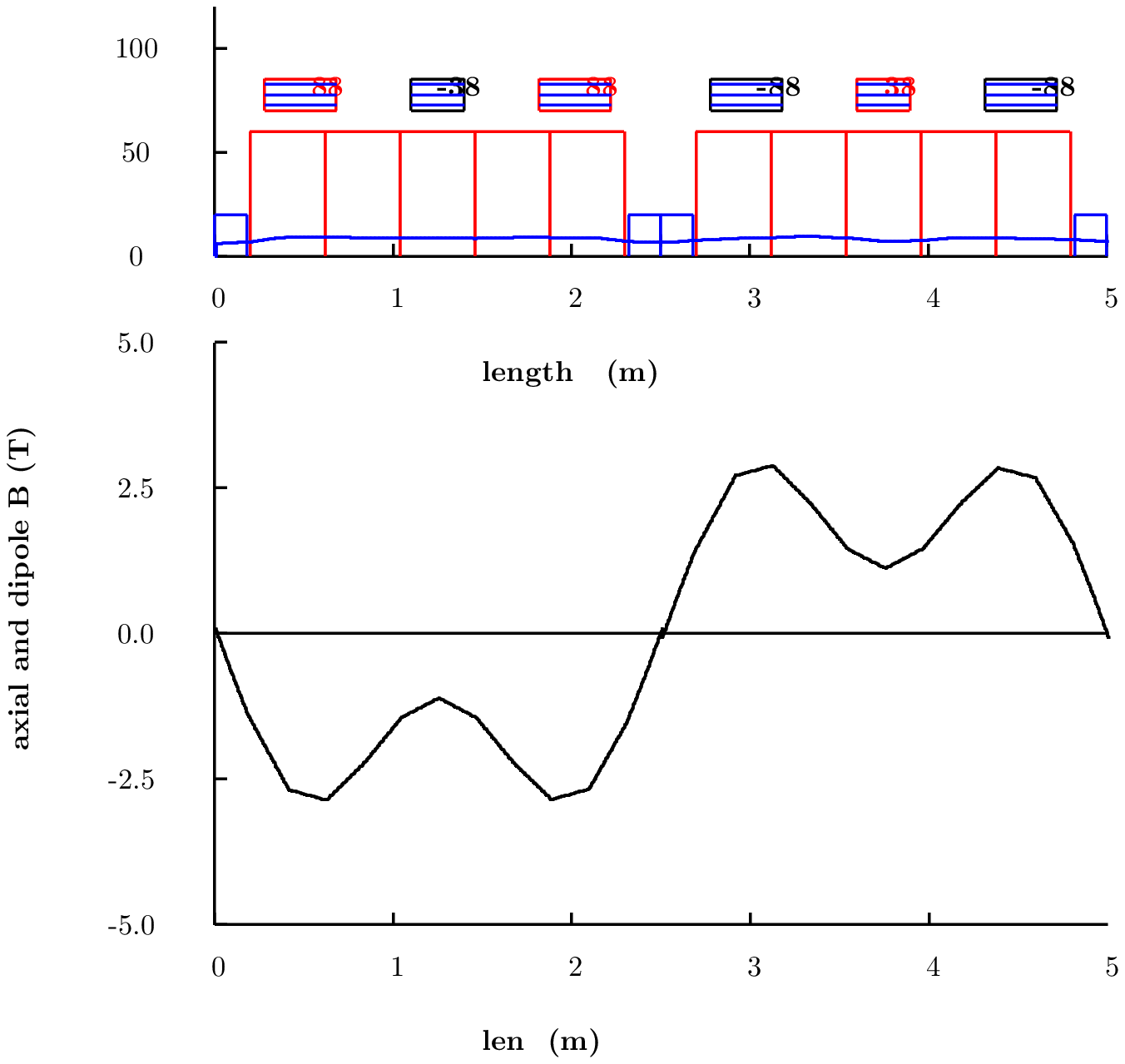}
\parbox{5.5in} % change 5.5in to \hsize for full-width caption
{\caption[ A cell of the cooling stage. ]
{\label{cool1} Top: half section through a super-FOFO cell of the cooling
apparatus, showing the
coil configurations, rf cells, and hydrogen absorbers.  Bottom: the axial
magnetic field \vs\ position.
}}
\end{center}
\end{figure}

\begin{figure}[htp]  % h = here, t = top, b = bottom, p = new page
\begin{center}
\includegraphics[width=6in, angle=0, clip]{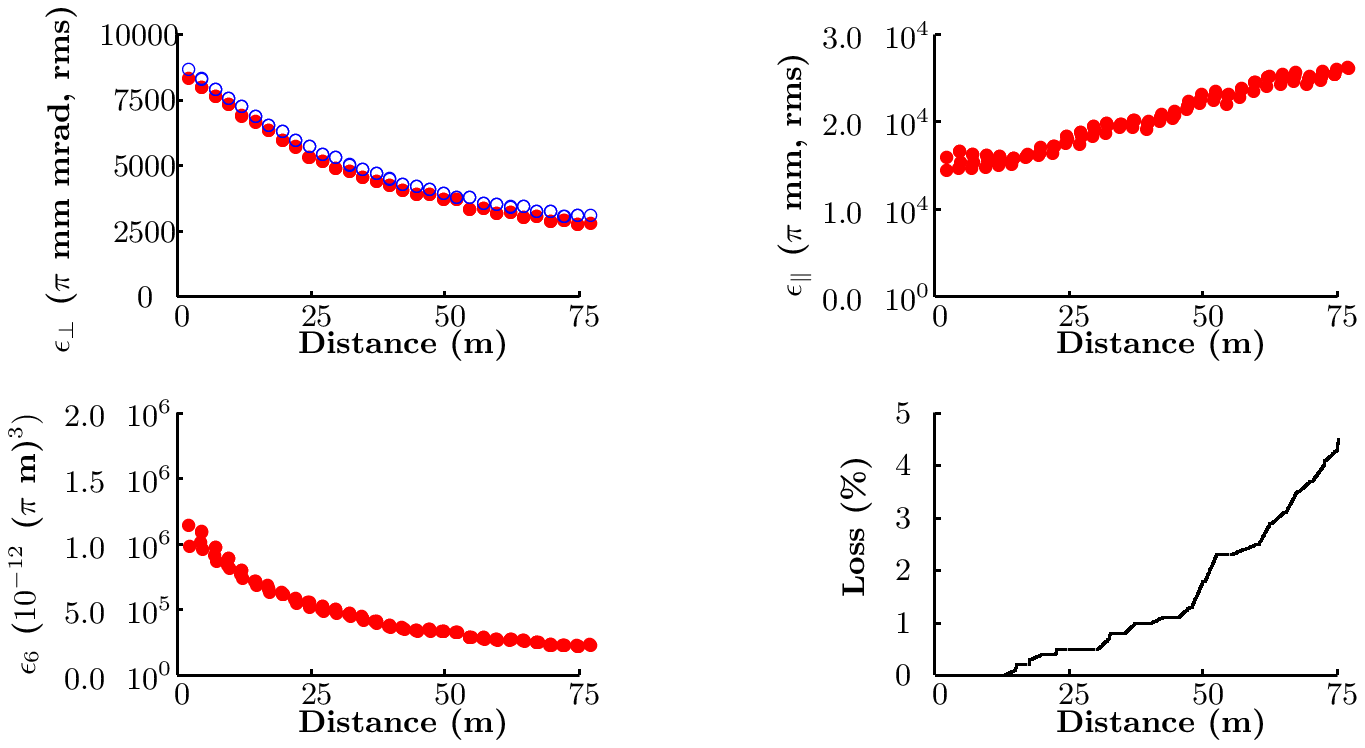}
\parbox{5.5in} % change 5.5in to \hsize for full-width caption
{\caption[ Cooling performance. ]
{\label{cool3} Transverse emittance (top left), 
longitudinal emittance (top right),
6-d emittance (bottom left), and particle loss (bottom right) \vs\
position during cooling.  
}}
\end{center}
\end{figure}

A more reasonable acceleration scheme considers 
an rms transverse acceptance of $\approx 1.5$ 
$\pi$ mm-radians. A cooling scenario based on the so-called super-FOFO
\cite{superfofo}
lattice of confining magnets (Fig.~\ref{cool1}) is under study.
The current simulation, using a fixed lattice and operating at a central
momentum of $185$ MeV/$c$ cools to below 3 
$\pi$ mm-radians, as shown in Fig.~\ref{cool3}.  It
does not achieve the required 1.5 $\pi$ mm-radians
because of Coulomb scattering at the end. Other lattices, 
with stronger fields easily reach the required final emittance, but do 
not accept the full initial emittance.  More work is needed here.

%\begin{figure}[htp]  % h = here, t = top, b = bottom, p = new page
%\begin{center}
%\includegraphics[width=4in, angle=0, clip]{palmer:cool2.eps}
%\parbox{5.5in} % change 5.5in to \hsize for full-width caption
%{\caption[ Momentum spread and bunch length during cooling. ]
%{\label{cool2} Momentum spread (top) and bunch length (bottom)\vs\ position
% during cooling. ]
%}}
%\end{center}
%\end{figure}

In a bunched beam, particles with large transverse amplitude must have
higher total velocity (higher energy) so that their longitudinal velocity,
$v_z$, remains matched to that of the bunch.  This is not practical for
relativistic beams, but can be arranged for nonrelativistic beams such as
considered here.

If the phase-rotation drift and buncher have a lattice with the same 
amplitude-$v_z$ properties as the cooling lattice, then the 
correlation is automatically generated.  Remember: the drift sorts 
particles by $v_z$, not energy. After the drift, 
their longitudinal position is a function of that $v_z$ which 
is the required correlated combination of energy and amplitude. The 
bunching, done in the same lattice (or one with the same properties) 
is also a bunching by $v_z$, not energy, so the correlation is 
preserved. And so into the cooling. 

Note that a simple solenoid will 
NOT do for the drift or bunching, since $v_z$ is a function 
not only of amplitude, but also of angular momentum. A solenoid of
 one sign gives a higher $v_z$ for one angular momentum sign 
than the other. Both drift and bunching must  done with alternating 
fields of some kind that maintain the canonical angular momentum near zero.
The super-FOFO lattice satisfies this requirement.

\subsection{Acceleration}

Acceleration of the cooled muon beam 
from 185 MeV/$c$ ($\approx$ 100 MeV) to 50 GeV is 
achieved by a linac followed by two recirculating linear accelerators (RLA's).

%The following parameters are taken directly from preliminary 
%work by Eberhard Keil for the Lyon Nufac99 workshop. They 
%will be updated as better numbers are available.

The present assumption is that the larger second (and possibly 
also the first) recirculating accelerator uses LEP superconducting 
cavities, or cavities with the same parameters and dimensions. 
The use of these cavities sets constraints on the minimum energy 
for which the required emittance can be transported. 
If the full ($\approx 10$ m long) cryostats, containing four 
cavities, are used as is, then this minimum energy is approximately 
8 GeV. This is taken as the approximate injection energy 
into the second RLA. If the cavities are rehoused 
individually in new cryostats, then the minimum energy is 
approximately 2 GeV. This is used as the approximate 
injection energy the first RLA.

More detailed considerations of the RLA's, and of the storage ring lattice,
are given in \cite{PJK}.

\subsection{Storage Ring}

{\bf Geometries}

\medskip

\begin{figure}[htp]  % h = here, t = top, b = bottom, p = new page
\begin{center}
\includegraphics[width=6in, angle=0, clip]{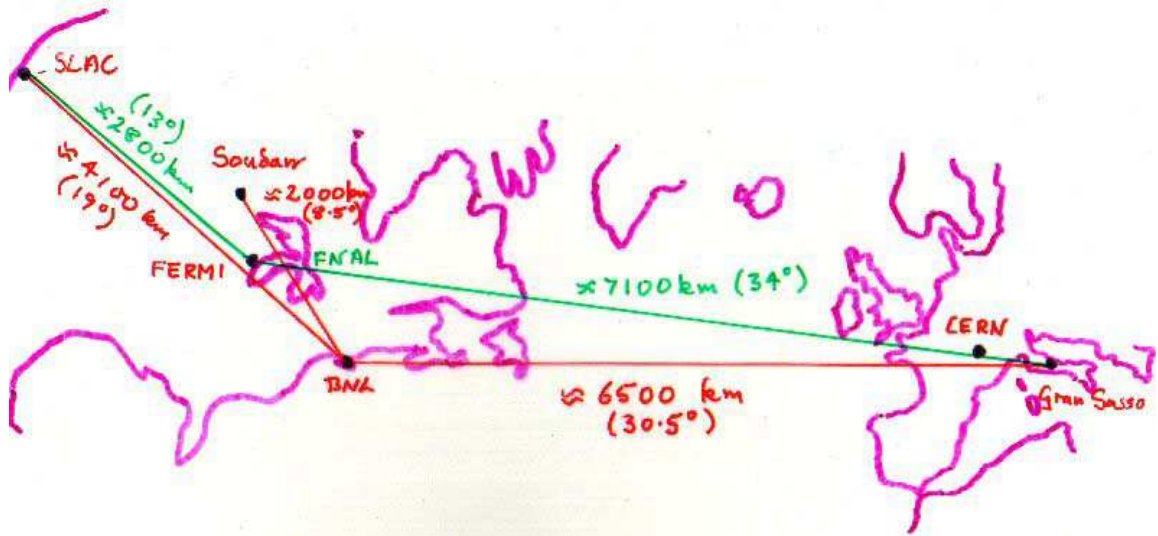}
\parbox{5.5in} % change 5.5in to \hsize for full-width caption
{\caption[ Global view of neutrino beam options. ]
{\label{map3} Neutrino beam paths between various possible sites for
source and detectors.
}}
\end{center}
\end{figure}

The geometry of the storage ring is site specific, being a 
function of both the ring and detector locations.
Figure \ref{map3} shows directions and direct distances from 
rings at BNL or FNAL to Gran Sasso, Soudan, and SLAC.
The circumference of such rings for 50 GeV muons must be of order 1 km,
even using bend magnets of several Tesla, so that a large fraction of the
length can be in neutrino-beam-producing straight sections.

% Sample LaTex Figure.  Refer to this figure in the text at fig.~\ref{ NAME }

\begin{figure}[htp]  % h = here, t = top, b = bottom, p = new page
\begin{center}
\includegraphics[width=4in, angle=0, clip]{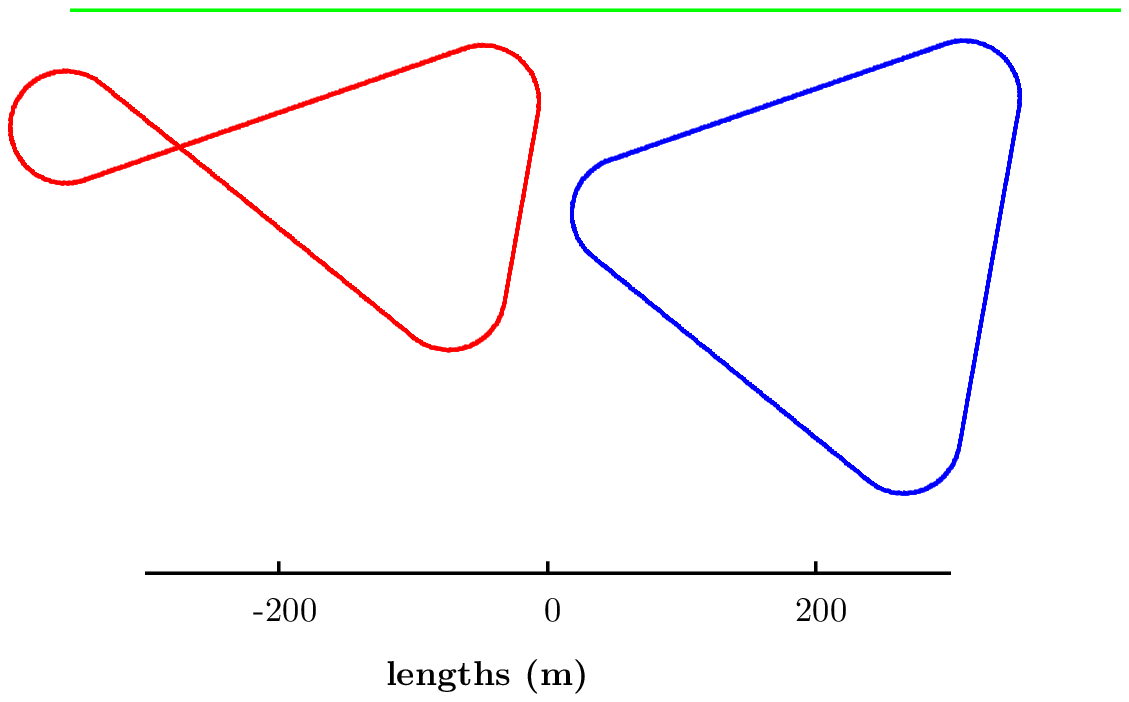}
\parbox{5.5in} % change 5.5in to \hsize for full-width caption
{\caption[ Storage ring geometries. ]
{\label{bowtie} Possible bowtie and triangular geometries for a muon storage
ring designed to deliver neutrino beams to two distant detectors.
}}
\end{center}
\end{figure}

For physics reasons (to separate MSW from vacuum oscillations), 
two differing ring to detector distances are required. If the 
two detectors lie in approximately opposite directions from the
 ring then it seems reasonable to design the ring with long 
sides that point to the two detectors, adding, if needed, a 
third straight to close the ring. Two  geometries are of particular interest
(Fig.~\ref{bowtie}):
\bi
\item A triangular geometry lying in a tilted plane. This 
minimizes the amount of bending required and maximizes the 
total straight for a given circumference. But, 
the lengths of the straights pointing at the 
two distant detectors is NOT maximized.
\item A ``bowtie", or figure-of-eight geometry, also lying 
in a tilted plane. This geometry uses more total bending, but does
maximize the important straights. It also has the interesting 
feature of not precessing the muon spins. A variant of the bowtie
 looks much the same but does not lie in a plane, so that there 
is a significant separation of the beams where they cross. 
In this case there is a slow precession of spin.

The bowtie can be made asymmetric so as to maximize 
the length of the upward straight.
\ei

To send a neutrino beam to a detector on another continent, a straight 
section in the storage ring must have angle at least $30^\circ$ to the
horizontal.  The resulting vertical extent of the storage ring is at least
100 m.  If the ring is below the surface, various geological issues must be
addressed.  It may be more practical to build the ring above ground and
bury it under an artificial hill.

\medskip

{\bf Lattice}

\medskip

The emittance that the storage ring must accept is estimated by supposing
there is 20\% emittance growth  in the accelerator in each of 
three directions.  We require an acceptance of 3 $\sigma$ in each 
of the 6 dimensions. 
If the bunch spacing is 1.7 m (corresponding to the 175 MHz bunching 
used here), then a reasonable maximum rms bunch length in the collider
 would be 6 cm. Thus the minimum momentum spread in this case would be 
$\sigma_p = 0.1\%.$

The rms beam divergence in the straight sections should be 
$\approx 0.1/\gamma$ in order a) to maximize the 
dependence of the $\nu_\mu$ to $\nu_e$ ratio on the polarization; 
and b) to assure that the flux observed is not significantly affected 
by the exact magnitude of this divergence.  To achieve this low 
divergence, the required beta function in the major straights is
$\beta_{\rm major~ straights}~\ge~75\ {\rm m}$.
For the up-going straight, aimed at a near detector on the surface, 
there is probably not such a stringent requirement on the beam 
divergence.  If the divergence here is required to be below 1/3 of 
$1/\gamma$, then
$\beta_{\rm upgoing~ straight}~\ge~8\ {\rm m}$.

\subsection{A First Look at Event Rates}

The numbers of surviving muons, per incident proton, at various stages of the
accelerator complex are summarized in Table~\ref{ptable15}.

\begin{table}[htbp] % h = here, t = top, b = bottom, p = on a new page
\begin{center}
\parbox{5.5in}  % replace 5.5in by \hsize if want full-width caption
{\caption[ Numbers of surviving muons.  ]
{\label{ptable15} The numbers of surviving muons after various stages in the
accelerator complex.
}}
\vskip6pt
\begin{tabular}{|l|c|cc|}
\hline\hline
$p$ driver energy (GeV)   &        &   24  & 16 \\
\hline
                        & Factor & $\mu$/p& $\mu$/p\\
\hline
Pions after Match ($<$ 1 GeV, forward)& & 0.66 &0.44 \s
After Phase Rotation \#1 (selected) &0.45 & 0.3& .2\s
After Phase Rotation \#2 (selected)&0.7 & 0.21&.14\s
After RF Capture        & 0.7 & 0.15&.1\s
After Cooling           & 0.9 &0.13 &.09\s
After Acceleration      & 0.7 & 0.092&.061\\
\hline
$n_\mu / (n_p~E_p)$~~~(GeV$^{-1})$     &    &.0038 & .0038\\ 
\hline\hline
\end{tabular}
\end{center}
\end{table}

The number of neutrino interactions per unit mass of a detector at distance
$L$ from a muon storage ring operating at energy $E_\mu$ scales as
\begin{equation}
N_{\rm events} ~\propto ~ N_\mu~E_\mu^3~L^{-2}.
\label{p6}
\end{equation}
For a proton power of 1.5 MW, and the muon survival efficiencies given in
Table~\ref{ptable15}, we would, in a
year of $10^{7}$~s of operation, obtain 4 $\times~10^{20}$ muons
decaying in the storage ring. If we take the fraction of the ring 
pointing to a given detector to be 0.25 (approximately as in the 
bowtie geometry), then the number of decays pointing to the 
given detector will be approximately $10^{20}$.

Table \ref{l_tab1} gives charged current neutrino interaction rates 
per kton-year
as a function of baseline length $L$ 
for an $E_\mu = 50$~GeV muon storage ring in which there are 
$1 \times 10^{20}$ unpolarized muon decays per year within a 
neutrino beam-forming straight section \cite{muc0051}. The rates are listed 
for 
\begin{enumerate}
\item[(a)]
 $\nu_e \rightarrow \nu_\mu$ oscillations with 
$\Delta m^2_{23} = 3.5 \times 10^{-3}$~eV$^2$/c$^4$ and 
$\sin^2 2\theta_{23} = 0.1$, 
\item[(b)]
 $\nu_e \rightarrow \nu_\mu$ oscillations with 
$\Delta m^2_{23} = 1 \times 10^{-4}$~eV$^2$/c$^4$ and 
$\sin^2 2\theta_{23} = 1$, 
\item[(c)]
 $\nu_e \rightarrow \nu_\tau$ oscillations with 
$\Delta m^2_{23} = 3.5 \times 10^{-3}$~eV$^2$/c$^4$ and 
$\sin^2 2\theta_{23} = 0.1$,
\item[(d)]
  $\nu_\mu \rightarrow \nu_\tau$ oscillations with 
$\Delta m^2_{23} = 3.5 \times 10^{-3}$~eV$^2$/c$^4$ and 
$\sin^2 2\theta_{23} = 1$. 
\end{enumerate}
Also listed are the rates for the unoscillated neutrino interactions, 
the corresponding statistical significance of the disappearance signal 
(numbers in parentheses), 
and the rates for the antineutrino interactions.

\begin{table}[htbp] % h = here, t = top, b = bottom, p = on a new page
\begin{center}
\parbox{5.5in}  % replace 5.5in by \hsize if want full-width caption
{\caption[ Neutrino interaction rates at a neutrino factory. ]
{\label{l_tab1} Neutrino interaction rates per kton-year at a neutrino factory
for four cases of neutrino-mass parameters as given in the text.
}}
\vskip6pt
\begin{tabular}{|ccc|ccc|ccc|}
\hline\hline
   & & Source & BNL & BNL & BNL & FNAL & FNAL & FNAL \s
   & & Detector & G.~Sasso  & SLAC & Soudan & G.~Sasso & SLAC & Soudan \s
 & & $L$ (km) &6528 & 4139 & 1712 & 7332 & 2899 & 732 \\
\hline
& Case & Mode &&&&&  &  \\
\hline
$\mu^+$ & (a) & $\nu_e \rightarrow \nu_\mu$ &
   90 & 160 & 190 & 63 & 180 & 200 \s
 & & $\nu_e \rightarrow \nu_e$ &
   1400 & 3600 & 16000 & 1100 & 8000 & $1.2 \times 10^5$ \s
 & & &
   ($2.4\sigma$)&($2.7\sigma$)&($1.5\sigma$)&
   ($1.9\sigma$)&($2.0\sigma$)&($0.6\sigma$) \s
 & & $\overline{\nu}_\mu \rightarrow \overline{\nu}_\mu$ &
   890 & 2200 & 9300 & 700 & 4800 & $7.0 \times 10^4$ \\
\hline
$\mu^+$ & (b) & $\nu_e \rightarrow \nu_\mu$ &
   $5 \times 10^{-2}$ & 0.86 & 1.5 & $3 \times 10^{-5}$ & 1.3 & 1.6 \s
 & & $\nu_e \rightarrow \nu_e$ &
   1500 & 3800 & 16000 & 1200 & 8200 & $1.2 \times 10^5$ \s
 & & &
   ($2.4\sigma$)&($2.7\sigma$)&($1.5\sigma$)&
   ($1.9\sigma$)&($2.0\sigma$)&($0.6\sigma$) \s
 & & $\overline{\nu}_\mu \rightarrow \overline{\nu}_\mu$ &
   890 & 2200 & 9400 & 700 & 4800 & $7.0 \times 10^4$ \\
\hline
$\mu^+$ & (c) & $\nu_e \rightarrow \nu_\tau$ &
   31 & 60 & 70 & 20 & 67 & 73 \s
 & & $\nu_e \rightarrow \nu_e$ &
   1400 & 3700 & $1.6 \times 10^4$ & 1100 & 8000 & $1.2 \times 10^5$ \s
 & & &
   ($2.4\sigma$)&($2.7\sigma$)&($1.5\sigma$)&
   ($1.9\sigma$)&($2.0\sigma$)&($0.6\sigma$) \s
 & & $\overline{\nu}_\mu \rightarrow \overline{\nu}_\mu$ &
   890 & 2200 & 9400 & 700 & 4800 & $7.0 \times 10^4$ \\
\hline
$\mu^-$ & (d) & $\nu_\mu \rightarrow \nu_\tau$ &
   450 & 570 & 650 & 410 & 620 & 680 \s
 & & $\nu_\mu \rightarrow \nu_\mu$ &
   760 & 3100 & $1.7 \times 10^4$ & 490 & 8000 & $1.4 \times 10^5$\s
 & & &
   ($35\sigma$)&($23\sigma$)&($12\sigma$)&
   ($40\sigma$)&($16\sigma$)&($4.6\sigma$) \s
 & & $\overline{\nu}_e \rightarrow \overline{\nu}_e$ &
   770 & 1900 & 8100 & 600 & 4100 & $6.1 \times 10^4$\\
\hline\hline
\end{tabular}
\end{center}
\end{table}

For comparison, the approximate numbers of events in the proposed 
CERN - Gran Sasso experiment (NGS) \cite{NGS}, and Minos \cite{Minos}
experiments, are 
given in Table~\ref{l_tab2}. It is seen that the numbers of events with the 
1.5-MW neutrino factory, in a detector at the same 730 km, is approximately 
100 times that in the NGS, or about 40 times the highest energy 
Minos example. 
%If the proton driver were upgraded to 4~MW, the 
%factors are approximately 300 and 100.

\begin{table}[htbp] % h = here, t = top, b = bottom, p = on a new page
\begin{center}
\parbox{5.5in}  % replace 5.5in by \hsize if want full-width caption
{\caption[ Comparison of neutrino interaction rates with Minos and NGS.  ]
{\label{l_tab2} Comparison of neutrino interaction rates per kton-year with 
Minos and NGS for beam conditions and neutrino mixing parameters as in
Table~\ref{l_tab1}. % {\sl These need checking.}
}}
\vskip6pt
\begin{tabular}{|l|c|c|ccc|}
\hline\hline
 & $\nu$ Factory & CERN-NGS && FNAL Minos & \\
\hline
$\ave{E_\nu}$ (GeV)  & 40 & 26 & 3 & 6 & 12 \\
$L$ (km) & 730 & 730 & & 730 & \\
\hline
$\nu_\mu \rightarrow \nu_\tau \rightarrow \tau$ & 680  & $\approx 7$
& $\approx 0$ & $\approx 30$ & $\approx 40$ \\
$\overline{\nu}_\mu \rightarrow \overline{\nu}_\mu$&140k  
& 1.5k & 0.46k & 1.4k & 3.2k \\
\hline\hline
\end{tabular}
\end{center}
\end{table}

%\clearpage

%\input nsfcollide.tex
\section{Muon Colliders}

A neutrino factory based on a muon storage ring is a possible first step 
towards a muon collider \cite{status}.  
This section briefly reviews the motivation for 
muon colliders, and sketches a sequence of such colliders.
%The physics case for muon colliders is further discussed in sec.~5.

The Standard Model of electroweak and strong interactions
has passed precision experimental tests at the highest 
energy scale accessible today.
Theoretical arguments indicate that new physics 
\textit{beyond  the Standard Model} associated with the 
electroweak gauge symmetry breaking and fermion mass
generation will emerge in parton collisions at or
approaching the TeV energy scale. It is likely that both hadron-hadron  and 
lepton-antilepton 
colliders will be required to discover and make precision measurements of the
new phenomena.  

The next big step  forward in advancing the hadron-hadron
collider energy frontier will be  provided by the CERN Large Hadron Collider
(LHC), a proton-proton collider with a center-of-mass (CoM) energy of 14~TeV 
 which is due to come into operation in the latter half of the next decade.

The route towards TeV-scale lepton-antilepton colliders is less clear. The 
lepton-antilepton colliders built so far have been $e^+e^-$ colliders, such as 
the Large Electron Positron collider (LEP) at CERN and the Stanford Linear  
Collider (SLC) at SLAC. In a circular ring such as LEP the  energy lost per 
revolution in keV is
$88.5\times E^4 / \rho,$ where the electron energy $E$ is in GeV, and the 
radius of
the orbit $\rho$ is in meters. Hence, the energy loss grows rapidly as $E$
increases. This limits the center-of-mass energy that  would be achievable in a
LEP-like collider. The problem can be  avoided by building a linear machine
(the SLC is partially linear), but with current technologies, such a        
machine must be very long (30-40~km) to attain the TeV energy scale.
Even so, radiation during the beam-beam interaction (beamstrahlung) limits
the precision of the CoM energy \cite{Tigner92}.

\begin{figure}[htp]  % h = here, t = top, b = bottom, p = new page
\begin{center}
\includegraphics[width=6in, angle=0, clip]{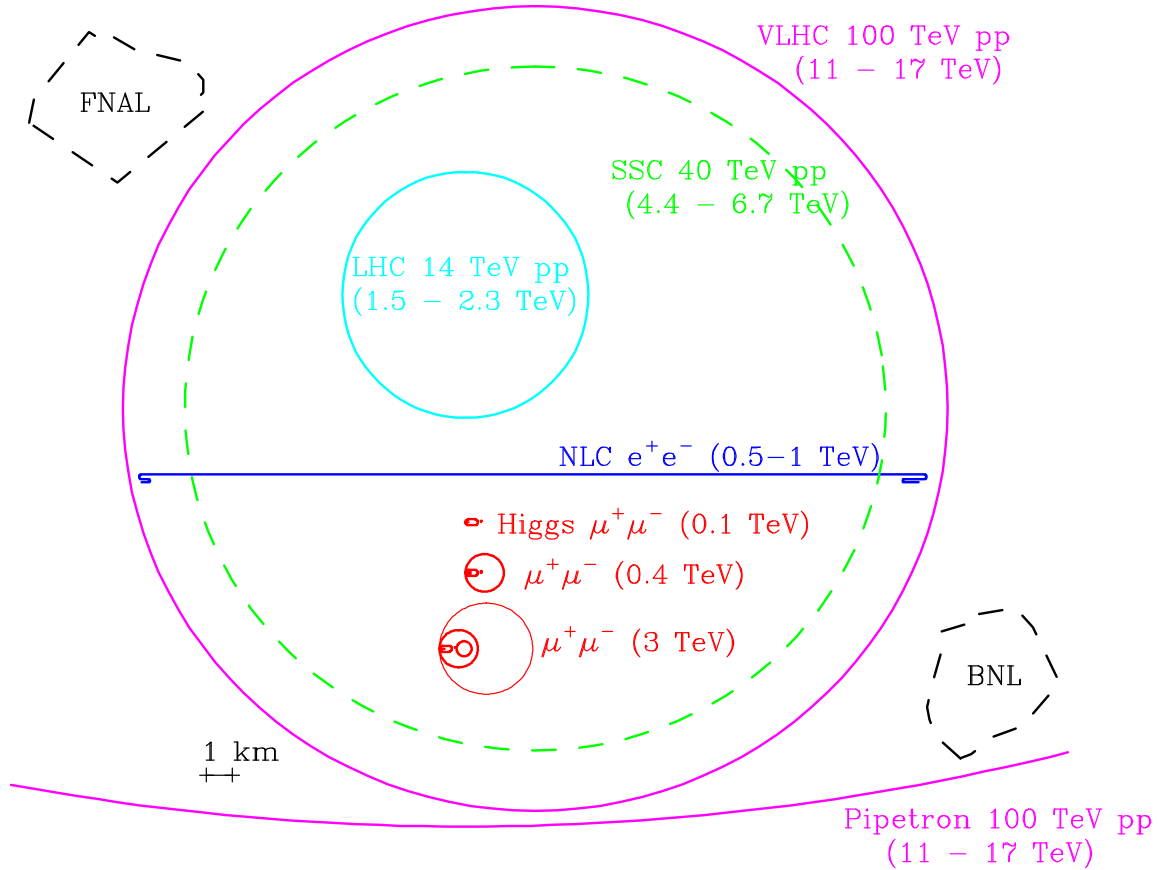}
\parbox{5.5in} % change 5.5in to \hsize for full-width caption
{\caption[ Various proposed high energy colliders.  ]
{\label{compare} Various proposed high energy colliders 
compared with the 
FNAL and BNL sites. The energies in parentheses give for lepton colliders their
 CoM energies and for hadron colliders the approximate range of CoM energies 
 attainable for hard parton-parton collisions.
}}
\end{center}
\end{figure}

\begin{figure}[htp]  % h = here, t = top, b = bottom, p = new page
\begin{center}
\includegraphics[width=5in, angle=0, clip]{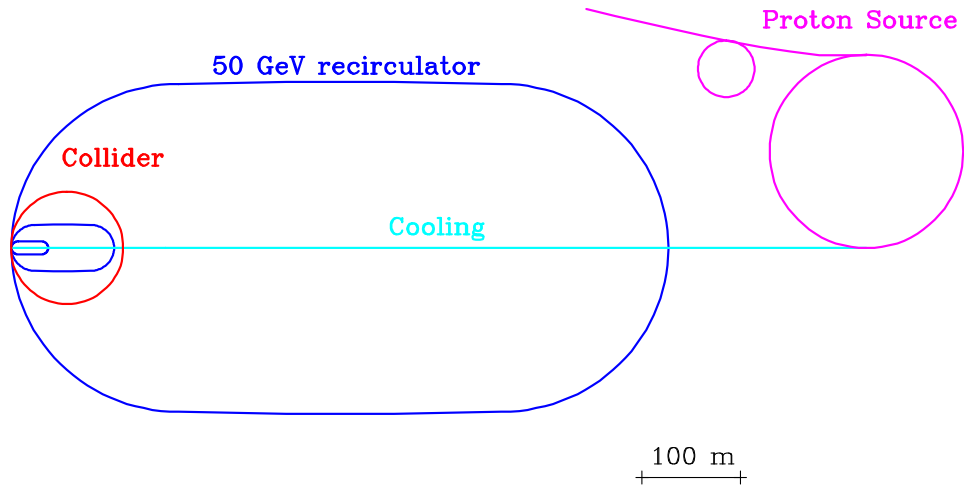}
\parbox{5.5in} % change 5.5in to \hsize for full-width caption
{\caption[ Plan of a 0.1-TeV-CoM muon collider. ]
{\label{plan1} Plan of a 0.1-TeV-CoM muon collider.
}}
\end{center}
\end{figure}

\begin{figure}[htp]  % h = here, t = top, b = bottom, p = new page
\begin{center}
\includegraphics[width=5in, angle=0, clip]{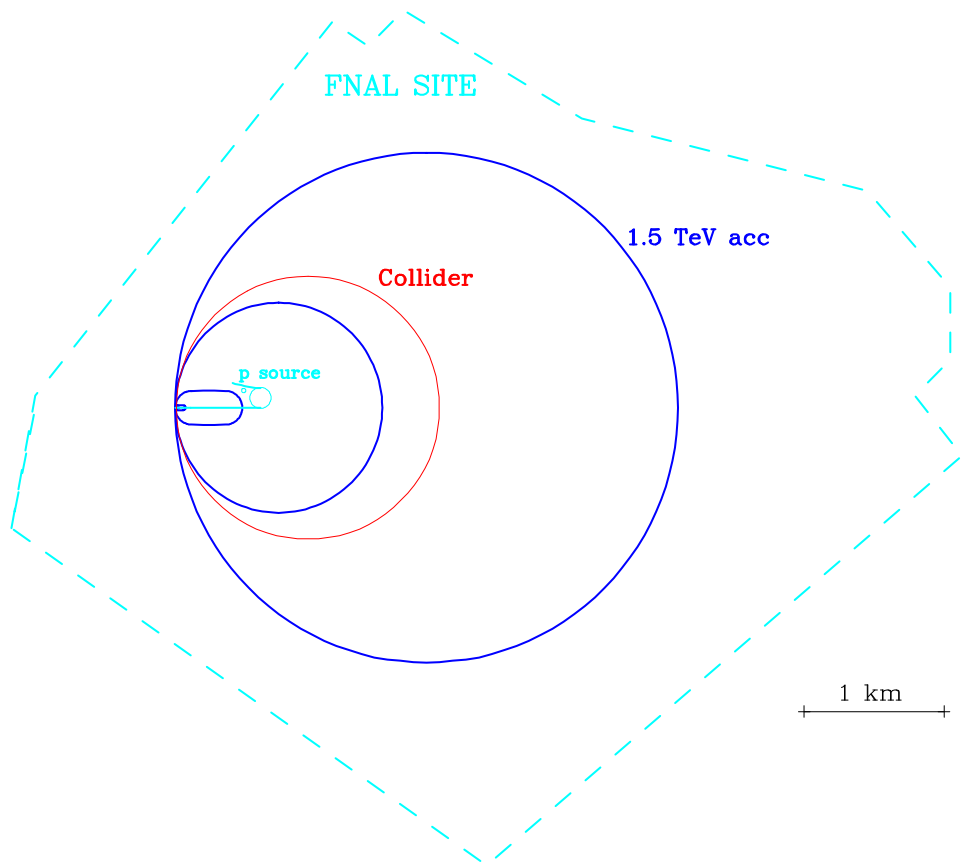}
\parbox{5.5in} % change 5.5in to \hsize for full-width caption
{\caption[ Plan of a 3-TeV-CoM muon collider.  ]
{\label{plan2} Plan of a 3-TeV-CoM muon collider shown on the Fermi National
Laboratory site as an example.
}}
\end{center}
\end{figure}

For a lepton with mass $m$ the radiative energy losses 
are inversely proportional to $m^4$. Hence, the energy-loss  problem can be
solved by using heavy leptons. In practice this means using  muons, which have
a mass $\approx 207$ times that of an electron. The resulting reduction in
radiative losses enables higher energies to be reached and  smaller collider
rings to be used \cite{Budker}. 
Estimated sizes of the accelerator  complexes required for
0.1-TeV, 0.5-TeV and 3-TeV muon colliders  \cite{status,colliders}
are compared with  the sizes of other
possible future colliders,  and with the FNAL and BNL sites in 
Fig.~\ref{compare}. Note that muon colliders with CoM energies up to $\approx 4$
~TeV would fit on these existing
laboratory sites. 
Figs.~\ref{plan1} and \ref{plan2} show possible outlines of the  0.1~TeV and 
3~TeV  machines. 
Parameters for 10 to 100~TeV colliders
have also been discussed \cite{100TeV}.

Muon  colliders offer significant physics advantages. The small
radiative losses  permit very small beam-energy spreads to be achieved.  For
example, momentum spreads as low as $\Delta P/P = 0.003\%$ are believed  to be
possible for a low-energy collider. By measuring the time-dependent decay
asymmetry resulting from the naturally polarized  muons, 
the beam energy could be determined with a precision of 
$\Delta E/E = 10^{-6}$ \cite{ref7}. 
The small beam-energy spread, together with the precise
energy determination, would facilitate measurements of the masses
and widths of any new resonant states scanned by the collider. In addition,
since the cross-section for producing a Higgs-like scalar particle in the
s-channel (direct lepton-antilepton annihilation) is proportional to $m^2$,
this extremely important process could be studied only at a muon collider and 
not at an $e^+e^-$ collider \cite{ref2b}.  And, of course, 
the decaying muons will produce copious quantities of neutrinos.  Even 
short straight sections in a muon-collider ring 
%not optimized for neutrino physics 
will result in neutrino
beams several orders of magnitude higher in intensity than presently
available, excellent for nonoscillation neutrino physics in a near detector.

\begin{figure}[htp]  % h = here, t = top, b = bottom, p = new page
\begin{center}
\parbox{3in}
{\includegraphics[width=3in]{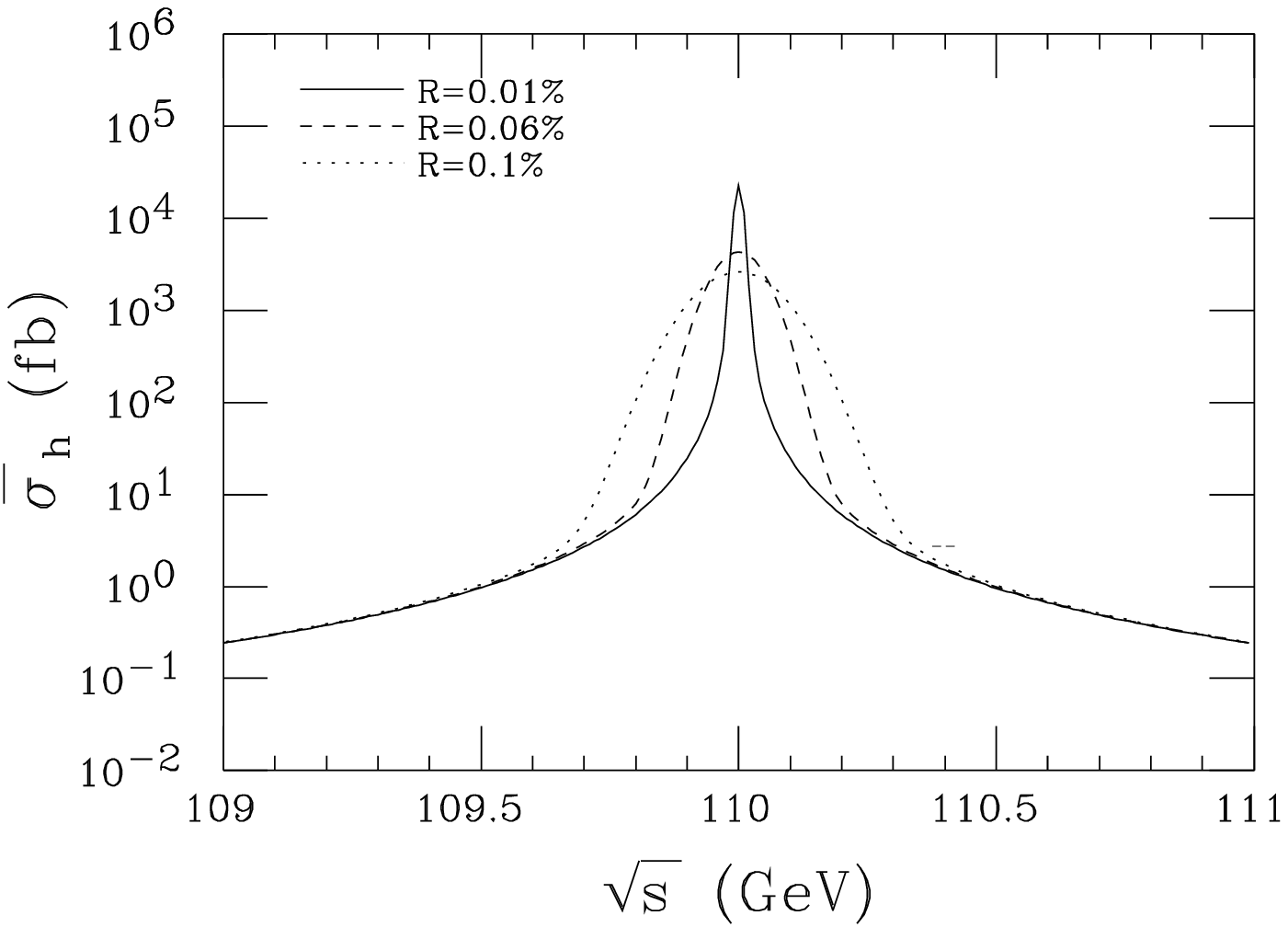}}
\parbox{3in}
{\includegraphics[width=3in, angle=0, clip]{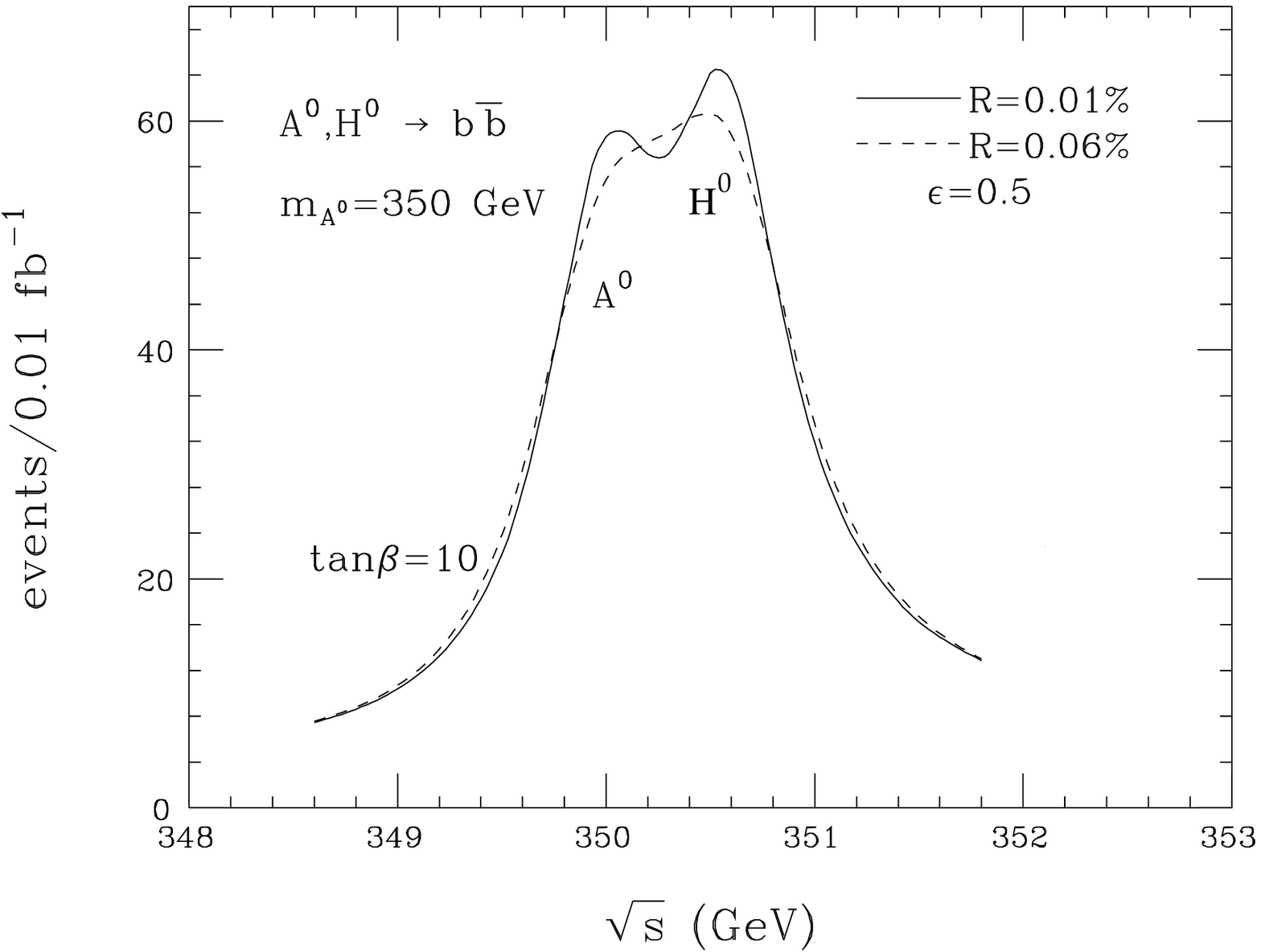}}
\parbox{5.5in} % change 5.5in to \hsize for full-width caption
{\caption[ Precision physics at a First Muon Collider. ]
{\label{s-chan-higgs} Left: effective $s$-channel Higgs cross section 
$\bar\sigma_h$ 
obtained by convoluting the Breit-Wigner resonance formula with a Gaussian 
distribution for resolution $R$. The mass of a light Higgs boson could be
determined to 1 MeV at a First Muon Collider.  
Right: separation of $A^0$ and $H^0$ signals for $\tan\beta=10$. 
From Ref.~\cite{Barger}.
}}
\end{center}
\end{figure}

The First Muon Collider will be a unique facility for neutral Higgs boson (or  
techni-resonance) studies through $s$-channel resonance production, as 
illustrated in Fig.~\ref{s-chan-higgs}.
Measurements can also be made of the threshold cross sections for 
production of $W^+W^-$,  
$t\bar t$, $Zh$, and pairs of supersymmetry particles --
$\chi_1^+\chi_1^-$, $\chi_2^0\chi_1^0$,  
$\tilde\ell^+\tilde\ell^-$ and $\tilde\nu\bar{\tilde\nu}$ -- that will  
determine the corresponding masses to high precision.
A $\mu^+\mu^-\to Z^0$ factory, utilizing the partial  
polarization of the muons, could allow significant improvements in  
$\sin^2\theta_{\rm w}$ precision and in $B$-mixing and CP-violating studies.

The Next Muon Collider
will be particularly valuable for reconstructing supersymmetric  
particles of high mass from 
their complex cascade decay chains. Also, any $Z'$  
resonances within the kinematic reach of the machine would give enormous  
event rates. The effects of virtual $Z'$ states would be detectable to high  
mass. If no Higgs bosons exist below $\sim$1~TeV, then the NMC would be the  
ideal machine for the study of strong $WW$ scattering at TeV energies.

The cost  of building a muon collider is not yet known. 
However, since muon colliders are relatively  small, they may be significantly 
less expensive than alternative machines. 

The front end of a muon collider is very similar to that of a neutrino
factory, with the important difference that the muon phase volume must be
cooled by a factor of $10^6$ rather than $\approx 100$.  During this larger
cooling, the longitudinal phase volume must shrink along with the transverse.
Since ionization cooling as proposed here directly cools only transverse
space, a muon collider must include an exchange between longitudinal and
transverse phase volumes so that cooling of the latter effectively results in
cooling of the former as well.

Another
 difference between the two machines is that a muon collider must provide 
muon bunches of both
signs simultaneously, while in a neutrino factory only one sign of muons is
utilized at any given time.  Further, a storage ring with long straight
sections optimized for neutrino beams is not ideal for high-luminosity
muon-muon collisions, particularly at lower energies.

\section{Research and Development}

\subsection{Historical Introduction}

The interest of the present proponents has evolved from our investigations of
muon colliders, the concept of which was introduced by 
Budker \cite{Budker}, and developed further by Skrinsky \etal\
 \cite{Skrinsky},  and by Neuffer \cite{Neuffer}. 
This work pointed out the significant challenges in
designing an accelerator complex that can  make, accelerate, and collide
$\mu^+$ and $\mu^-$ bunches all within the muon lifetime of $2.2\,\mu$s 
($c\tau=659$~m), and provided preliminary sketches of technical solutions.

A concerted study of a muon collider design has been underway  since 1992 
\cite{mushops}.
By the Sausalito
workshop \cite{ref3b} in 1995 it was  realized that with new ideas and modern
technology, it may be feasible to make muon bunches containing a few  times
$10^{12}$ muons, compress their phase space and  accelerate them 
%the resulting still intense bunches 
up to the multi-TeV energy scale before more than about 3/4 of them have
decayed. With careful design of the collider ring and  shielding it  appears
 possible to
reduce to acceptable levels the backgrounds  within the detector that arise
from the very large flux of electrons produced  in muon decays. These
realizations led to  an intense activity,  which resulted in the muon-collider
feasibility study report \cite{ref6a} prepared for the 
1996 DPF/DPB Summer Study on High-Energy Physics (the Snowmass'96  
workshop). 

 Encouraged by further progress in developing the 
muon-collider concept, together with the growing interest and involvement of 
the high-energy-physics community, the Muon Collider Collaboration
became a formal entity in May of 1997 \cite{collabsite,charter}.  
An overview of the
activities and plans of the Muon Collider Collaboration is given in
\cite{status}.

That a neutrino factory would be a good first step towards a muon collider
has been explored in two Collaboration workshops \cite{mup,BNL98} as well
as by ECFA/ICFA study groups \cite{CERN9902,muonsatCERN}.  The NuFact'99
Workshop \cite{nufact99} in June 1999 provided a focus for international
interest in neutrino factories, motivated by the outstanding physics
prospects plus the need for truly global facilities for long baseline
neutrino physics.

Accordingly, the Muon Collider Collaboration has recently changed its name to
the Neutrino Factory and Muon Collider Collaboration, and is redirecting its
efforts towards an early realization of a neutrino factory.  
A Muon Steering Group \cite{MUG} has been formed in Europe to coordinate 
efforts there towards the same goal.  These two structures
are formally distinct, but there is excellent communication among members of
the two groups.

The Muon Collider Collaboration has proposed an R\&D program that features
hardware studies of two key aspects of a muon collider: 
\begin{itemize}
\item
Targetry, capture and phase rotation at a muon source \cite{targetprop},
\item
Final-stage ionization cooling at a muon collider \cite{MUCOOL},
\end{itemize} 
in addition to an ongoing program of machine theory and simulation.
There has been one outside review of the R\&D program \cite{MUTACpresent},
 conducted in July,
1999 by the Muon Technical Advisory Committee (MUTAC) of the Muon Collider
Oversight Group (MCOG).  The MUTAC report \cite{MUTACreport} and the
MCOG report \cite{MCOGreport} following this review emphasized that
the R\&D program should be conducted in the context of ``a more formal,
long range, R\&D plan'' with a ``focus on one object for a complete,
detailed study''.  They noted that ``a neutrino source appears as the most
likely possibility'' for that study.

%This document is part of the process of defining the longer range R\&D program
%for a neutrino factory.

Consistent with the emerging emphasis on a neutrino factory, the Collaboration
is re-examining its R\&D priorities, as well as seeking broader support for
these activities.

\subsection{R\&D Needs for a Neutrino Factory}

The overall path of an R\&D program involves conceptual design, demonstration
of feasibility of novel components, followed by cost optimization.  A neutrino
factory based on a muon storage ring is still very much in the early phases of 
conceptual design, with some items identified as needing verification as to 
their feasibility.  Nonetheless, there are some pressures to concern ourselves 
with cost issues already at this early stage \cite{FNALsitestudy}.

The prominent R\&D issues for a neutrino factory are listed below, following
the sequence of components in the accelerator complex.
\begin{enumerate}
\item
{\bf Coherent design concept} of an entire neutrino factory.
\item
{\bf Proton driver}: 1-4 MW, 5-15 Hz, $\approx 5 \times 10^{13}$ protons per 
bunch, {\bf 1 ns bunch length}.   The critical issue of short bunch length
in a proton synchrotron is 
under study by an ANL-BNL-FNAL-KEK-LANL collaboration 
\cite{ref10,agsbunch,indinskek,PSR}.
\item
{\bf Pion yields from proton-nucleus collisions}.  
A neutrino factory would collect
very low energy pions, for which the rate is maximal.  Such pions are
partially absorbed in the targets of most prior production experiments, so the
data are questionable.  A recent measurement by members of the Collaboration
should improve our knowledge from proton beams of 6-24 GeV \cite{E910}.  An
experiment to study yields from 2-GeV protons is being considered at CERN
(sec.~6.4) in the context of the option for a proton driver linac.
\item
{\bf Production target}.  Proton pulses of 70-280 kJ energy and  1~ns length
are incident on the target, leading to substantial issues of ``shock'' damage,
cooling and materials survivability in a high radiation environment.
While it is natural to consider solid targets, their viability is considered
marginal, and liquid targets are the alternative.  For maximal pion production,
a free liquid jet target is to be preferred in principle.  There is no 
example of such a target.
\item
{\bf Capture solenoid}.  Optimal pion yields are obtained when the target is
surrounding by a solenoid of field $\approx 20$ T, followed by an adiabatic
transition to a solenoidal channel of a few T.  Such a magnet would be
a superconducting hybrid with a resistive insert \cite{nhmfl-lhfs}.
A key question is the effect of radiation damage on such a device.
\item
{\bf Beam dump}.  The 1-4 MW proton beam is dumped inside the target/solenoid
system.  A flowing liquid dump may be more appropriate than a solid dump.
\item
{\bf First Phase Rotation}.  If polarized muon beams are to be obtained, the
production target must be quickly followed by a high-gradient, low-frequency
rf system, combined with a solenoid channel, to bunch the pion/muon beam.
Little is known about the viability of such a system near an intense 
radiation source.
\item
{\bf Mini Cooling}.  The use of a passive liquid hydrogen absorber to provide
initial transverse cooling of the muon beam by a factor of two is well 
understood in principle,
although it never has been demonstrated.
\item
{\bf Second Phase Rotation}.  For the second step in the bunching process,
the muons must be accelerated by 80-100 MeV to restore the energy lost in the
mini cooling.  A large acceptance induction linac with a programmed waveform
is required.  The parameters of the linac are somewhat beyond those
presently demonstrated.
\item
{\bf Bunching to $\approx 400$ MHz}.  This is believed to be relatively
straightforward.
\item
{\bf Ionization Cooling}.  The challenges of further acceleration and storage 
of the muon beam will be substantially easier if the transverse phase area of
the beam can be reduced by an additional factor of 10.  This cannot be
accomplished in a single step of ionization cooling, but must involve
alternating ionization cooling and rf acceleration, all in a magnetic
channel.  This is a key area for study, and a hardware demonstration is very
appropriate.
\item
{\bf Acceleration}. The acceleration from $\approx 100$ MeV to $\approx 50$
GeV is best accomplished in recirculating linacs with superconducting rf
cavities.  Rather large acceptances are required, and the machine parameters
are again somewhat beyond those presently demonstrated.
\item
{\bf Muon Storage Ring}.  The desire for multiply directed neutrino beams
with very small angular divergence leads to novel designs for the storage
ring, whose plane is far from horizontal.  Besides issues of lattice design,
there will be considerable civil engineering challenges in building such a
ring.  
\end{enumerate}

The R\&D needs for a muon collider are very similar, but with additional
challenges in cooling and storage ring design.  At least four orders of 
magnitude more cooling (including continual exchange between transverse and
longitudinal emittance) are required for a muon collider than a neutrino
factory, and a rather different ring is needed to maximize collider
luminosity than simply to hold the muons while they decay.

A sense of the Collaboration's views as to the relative urgency of 
addressing the above issues is given by the following ranking.  Given in
parentheses are the institutions presently involved in R\&D into these
topics.
\begin{enumerate}
\item
Coherent design study (the Collaboration as a whole).
\item
Target, dump, phase rotation (ANL, BNL, UCLA, CERN, LBNL, ORNL, Princeton).
\item
Ionization cooling (ANL, BNL, Budker Inst., UC Berkeley, UCLA, FNAL, IIT, 
Indiana U., LBNL, NHMFL, Northern Illinois U., Princeton).
\item
Induction linac (LBNL).
\item
Recirculating linacs, superconducting rf (Jefferson Lab).
\item
Storage ring design (BNL, CERN, FNAL, LBNL).
\item
RF power sources (BNL, CERN, FNAL, LBL + industry).
\item
Effects of radiation on superconducting magnets (MSU).
\item
Fabrication of superconducting magnets (LBNL, NHMFL + industry).
\item
RF bunching.
\item
Engineering of a tilted ring.
\item
Engineering of ``conventional'' facilities (FNAL, ORNL).
\end{enumerate}

Proton driver issues are very site specific, and have been left off the
second list as being somewhat 
outside the scope of the Neutrino Factory and Muon Collider Collaboration.
Pion production cross sections were also left off the second list as being
adequately addressed by efforts largely outside the Collaboration.

The strategy for pursuit of the R\&D topics listed above is an
interesting challenge in itself.  The variety of questions is large, and
several go beyond the scope high-energy accelerator experience.
A neutrino factory is still too novel a concept to be sponsored as a
well-defined program at a single accelerator laboratory.  The cooperative
efforts of people at many institutions is needed to bring the concept
of a neutrino factory to the stage of a formal Conceptual Design.
%Yet, considerable coordination of effort is required for R\&D to proceed in
%a timely manner.

The Neutrino Factory and Muon Collider Collaboration has taken responsibility
for the coordination of multi-institutional R\&D efforts on the 
non-site-specific aspects of a neutrino factory.  While the topic of
research is largely accelerator physics, the operation of the Collaboration
is more similar to that of a large experimental physics group proposing
a novel detector than to that of  past accelerator projects.  The
Collaboration has been successful in providing a means of groups
of people working together, as facilitated by numerous workshops
\cite{workshops}, video conferences \cite{video}, web sites
(see links at the primary site \cite{collabsite}, and
an archive of technical documents \cite{notes}.  Additional efforts are
needed to enhance the coherence of this work, an important step of which will
be the appointment of an R\&D Coordinator.  There remains the issue of
the response of the Collaboration to the advice of the MUTAC that ``the first
round of Design and Simulation activities may requires 10-15 accelerator
experts for 1-1/2 to 3 years.  The coherence required for success in this
activity demands full-time workers in close communication."

To carry out the R\&D program sketched above, the Collaboration seeks
additional resources in two categories:
\begin{enumerate}
\item
Support for a core group of physicists, most of whom are in residence at a
single site, likely a national laboratory.  Support is sought for both
staff positions, and for visitors who would locate at the core site for
at least several months at a time.
\item
Support for the various particular R\&D topics listed above, which work
may well be effectively pursued at diverse labs and universities.
\end{enumerate}
Partial support for Collaboration efforts in both categories of work is 
presently available via direct funds from the Advanced Technology R\&D
Program (and to a much smaller extent from the Physics Research Program)
of the U.S.\ Department of Energy, as well as from discretionary funds
at the major U.S. national laboratories.  
The state of Illinois has made a commitment to a university consortium
heading by IIT for funding beginning
in year 2000.  CERN is starting an R\&D program, (sec.~6.4), with initial
funding in the present fiscal year.

We have previously estimated that
a robust R\&D program for muon colliders would require about \$15M/year.
A very similar figure is appropriate for neutrino factory R\&D, as this
is effectively a transformed muon collider R\&D program.  Our present funding
is approximately 1/2 of this amount.

The favorable outcome of our R\&D program is, of course, the construction of a
neutrino factory.  Prior to this, we anticipate the elevation of effort to
that of a major project at at least one national laboratory.  The role of
the Collaboration will no doubt evolve significantly in such case, but it
can and should continue to play a key role in harnessing the diverse
resources needed to design a neutrino factory.  The original role of the
Collaboration as a vehicle for broad-based efforts towards a muon collider 
will again be important as a neutrino factory becomes associated with a
particular site.

%It is too early to estimate a definitive time scale for R$\&$D, 
%a time scale for construction and the 
%actual cost of a complex. 
%These can be expected to be a result of the initial R$\&$D within the next two
%years. It is however good to remind that the R$\&$D effort for muon colliders is
%estimated to 15 M\$ per
%year. 
%This figure may apply to neutrino factories as well but over a shorter time 
%since the tasks 
%to be achieved are essentially the same and the cooling requirements much 
%less severe.

\subsection{The Potential of Muon-Beam-Based Particle Physics
and the NSF-Supported Community}

%{\sl [M.~Tigner, Cornell University]}

Just 20 years ago the DoE assembled a HEPAP Subpanel on Accelerator R and D.  
In the letter conveying their report to HEPAP the Subpanel Chair wrote:
``You will note that in the 50 odd years of American accelerator science 
associated with particle physics research, enormous  strides in increasing 
particle beam energies and in decreasing unit costs have been and are being 
made. ... Our primary conclusion is that, despite the spectacular 
past and present accomplishments of the field, we must redouble our efforts 
to improve the cost effectiveness of our accelerators if the needs of 
US particle physics are to be met in the resource-limited situation in which 
we find ourselves..."

Unfortunately, as recent history and current events show, this observation 
is even more apt today than it was those 20 years ago.  This is not for 
want of zeal and good ideas.  In the intervening years considerable 
progress in understanding the fundamentals of ``classical" accelerator 
science and improving classical accelerator technology has been made. 
There have also been some advances based on technologies not previously 
used in elementary-particle-physics accelerator work, \eg, laser and 
plasma technology.  
It is, however,  a fact that none of these efforts, to date, have 
qualitatively changed the cost of providing significant luminosity 
at what is now the energy frontier.  Consequently, it is not an 
exaggeration to say that today we are in danger of pricing ourselves 
out of the market.

In recent years, as accelerator science and technology have become 
more and more sophisticated and thus more specialized, the task of 
developing the accelerators needed for the future has more and more 
been left to experts -- specialists in accelerators.  They have done 
an excellent job indeed. The capabilities of today's accelerators 
would have even been unthinkable 20 years ago.  

Nevertheless, we find ourselves in the unenviable position that each 
new energy-frontier facility being discussed turns out to be in the 
multi-billion dollar class.  This difficulty might find a direct 
political solution from time to time as history unfolds and the 
competitive juices flow strongly again.  However, if this had been 
the path followed in the past, elementary particle physics
 would not be able to ask the 
compelling questions that it can ask today.  Thus, the direct approach 
of tackling the problem scientifically and technologically is likely 
to be more dependable -- no guarantees.  One obvious avenue is to 
broaden the scientific and technical idea base which might support 
significant improvements in accelerator cost effectiveness.  
This implies that the problem, OUR problem, needs to be brought 
more directly and effectively to the stakeholders in elementary particle
physics,  that 
is to say, to the university and laboratory community of experimental 
and theoretical physicists who now concern themselves primarily with 
the particle physics and detector instrumentation.  This has been 
tried to some degree in the past, with only modest success.  Today 
the need is more apparent and, in addition, we now have a made-to-order 
challenge that needs all the new and non expert ideas that 
it can get - the possibility of doing elementary particle physics
with high energy muon 
beams through muon acceleration and storage for intense neutrino 
production, and later directly for $\mu$-$\mu$ collisions.

Many aspects of this concept are new enough that even the experts 
have to start from scratch.  This stems from the unusual requirement 
that the job has to be done quickly owing to the finite life of the 
beam and, perhaps more importantly, that an enormous spread in beam 
momenta and angle must be accommodated if the required capture 
efficiency is to be met.  The situation is somewhat analogous to the 
situation in accelerator science forty-odd years ago when folks 
tracked particles through magnetic fields using the Runge-Kutta 
method with a Marchant calculator.  All of that calculation with 
the attendant trial and error struggle to find workable system 
designs was made obsolete with the elegant theoretical work of 
Courant, Livingston and Snyder and many others.  They discovered 
powerful methods for dealing with paraxial ray beams of relatively 
narrow energy spread.  These methods are of limited use in studying 
the optics of a muon-based neutrino source or collider where 
nonlinearities are controlling rather than perturbations.  
Not only that, but the main optical components will probably 
have to be solenoids, a device which has heretofore not been 
used for the principal focusing and bending elements in high 
energy machines.  Trying to master all this puts everyone more 
or less on the same footing and begs for some new  tactics 
from the classical mechanics buffs among us.  Latter day 
Courants, Livingstons and Snyders are sorely needed.

There are yet other unprecedented challenges.  The science and 
technology of quickly reducing the phase space volume of the 
beam needs developing before muon beams of the required brightness 
can be produced.  While basic ideas for accomplishing this via 
ionization cooling have 
been around for years, the practical problem of realization is 
also new to the experts and involves very fundamental physics, 
some of which is not yet known with the depth required to 
support the needed technology.  

%More detail on these and other pressing scientific and technical 
%matters is found in the accompanying documents outlining the 
%physics potential and challenges to its realization as well 
%as possible facility concepts.

Considerable attention has been focused on the potential physics 
opportunities  for muon-based neutrino science and on possible 
means for attacking it.  A Neutrino Factory and Muon Collider 
Collaboration (NFMCC), formed of members from the DoE supported 
Labs, Budker Institute for Nuclear Physics and some universities, 
has been formed and has been at work for some time.  An idea of 
the progress that has been made is presented in secs.~1-5 of this
document, which makes clear that a resolution of the basic 
particle and accelerator physics issues remains in the future 
and that more ideas, more work and much R\&D lie ahead even in 
evaluating whether our community can and should propose such a facility.  

Taking into account the fact that there is a great deal of talent, 
knowledge and expertise in the university community -- both DoE and 
NSF supported -- not now engaged in addressing the pressing accelerator 
issues, it would seem most appropriate to try to tap that pool.    
To make this possible, two things at least are needed.  First, 
they have to be made aware of the possibilities and challenges.  
This the world community in general and the NFMCC in particular 
are doing.  The NFMCC will be emphasizing this aspect more in the 
coming year.  Second, modest start-up resources are necessary for 
preliminary engagement with the accelerator challenges, resources  
such as funding for post docs, some computing and modest beginnings 
of technical R\&D.  It is with respect to these needed monetary 
resources that we are addressing this 1999 MRE Panel.  

Within one to two years it may well become apparent that large 
R\&D expenditures, \ie, 10's of M\$,  by NSF-supported university 
groups working on the accelerator aspects of muon-beam elementary particle
physics, will 
be appropriate.  This would require a joint application for MRE 
funding.  The effort required to plan and justify such an application 
for review by the physics community, being an unusual enterprise, 
needs unusual support.  Our hope is that this MRE Panel will  
appreciate this special need and recommend to the NSF that, 
where possible, they provide start up resources for currently 
supported university particle physics groups to become so engaged 
in the knowledge that this work may well lead to an MRE proposal 
in the not too distant future.

\subsection{European R\&D Activities on Muon Storage Rings and Neutrino
Factories}

%{\sl [This section provided by the Steering Group of European Studies on Muon
%Storage Rings. \cite{MUG}.]}

There is growing interest in Europe for muon storage rings and particularly
neutrino factories.
Several working groups have been set up to study: 
\begin{enumerate}
\item
The accelerator aspects of a neutrino factory at CERN;
\item
The physics of neutrino oscillations; 
\item
The opportunities offered by high-intensity neutrino muon and hadron beams; 
\item
The physics opportunities of the extension of a neutrino factory
to a precision muon collider \cite{muonsatCERN}.
\end{enumerate}

%The following summarises the R\&D that are presently considered.

Discussions with physicists and accelerator engineers from European
institutes and laboratories, and from CERN, have focused on identifying
important missing elements in the currently debated designs of muon
storage rings, with a view to avoiding duplication of efforts while
contributing significantly towards the design of a neutrino factory.

The European community is considering the following R\&D projects:
\begin{enumerate}
\item
{\bf A hadron production experiment at the CERN-PS}.
The aim is to measure charged pion production by 2-16~GeV
protons, data that are needed for a quantitative design of pion capture
and phase rotation.  The very same experiment can be extended
to hadron production by pions, so as to deliver the entire set of data
that is needed for a reliable calculation of the atmospheric neutrino
flux.
\item
{\bf A large-angle muon scattering experiment.}
This experiment would measure with high precision the large-angle scattering
of muons with momentum of a few hundred MeV/$c$ in various materials
including
liquid hydrogen, as theoretical calculations are not reliable enough to
assess the performance of ionization cooling of muons.
\item
{\bf Exposure of an rf cavity to radiation and a magnetic field.}
One of the big unknowns is the reliability of operation of the rf cavities
which are currently discussed for pion capture and phase rotation, and
which will have to operate in a high-radiation field and possibly in
strong solenoidal magnetic fields. Experiments with pulsing rf
cavities would also be performed with a view to achieving higher gradients.
\item
{\bf High-power target tests.}
Current design work is focussed on targets which withstand a beam power
of 4 MW or even larger. While not considered impossible, this is a
daring goal for which, however, considerable know-how is available in
Europe (CERN, GSI, KFA Julich, PSI, RAL), which can and should be channeled
towards an interesting and forward-looking challenge.
\end{enumerate}

This proposed program of experimental R\&D work in Europe is by
and large complementary to the R\&D activities planned or under way in the USA.
This experimental work is augmented 
by theoretical studies, both in the area of physics and detectors,
and in the area of accelerator design (proton linac, fast-cycling
synchrotron, muon recirculators).

\section{Acknowledgements}

This document was largely assembled from existing sources, which have been
cited among the references.  Here we would like to identify and thank those
individuals who contributed paragraphs or more of the text.  Section 1 is
based in part on notes by R.~Shrock.  Sections 2 and 3 are adapted from
\cite{whitepaper}, which was edited by B.~Autin from contributions by
A.~Donini, M.B.~Gavela, P.~Hern\'andez, S.~Rigolin, and S.~Petcov 
(secs.\ 2.1 and 2.2), D.A.~Harris (sec.\ 2.3), and K.S.~McFarland
(sec.\ 3), among others. Section 4 is from R.B.~Palmer with additional
material from S.~Geer (sec.\ 4.10) as well as C.~Johnson and E.~Keil.
Sections 5 and 6.1 are adapted from the Muon Collider Status Report 
\cite{status}
which was edited by J.C.~Gallardo; the pieces used here are from V.~Barger,
S.~Geer, J.~Gunion, and R.B.~Palmer.
Section 6.2 is adapted from notes by A.~Sessler. Section 6.3 is by M.~Tigner.
Section 6.4 is by the Steering Group of European Studies on Muon
Storage Rings \cite{MUG}, chaired by A.~Blondel.

We especially thank those contributors who are not members of the Neutrino
Factory and Muon Collider Collaboration.

\clearpage

\end{document}